\documentclass[a4paper,12pt]{elsarticle} 

\usepackage{indentfirst}
\usepackage{anyfontsize}
\usepackage{subfig}
\usepackage[dvipsnames]{xcolor}

\usepackage{amssymb}
\usepackage{caption}
\usepackage{overpic}
\usepackage{amsmath}

 \usepackage{relsize}
  \usepackage{bm}
\usepackage[pagewise]{lineno}

\begin{document}
\begin{center}


{\bf Moment Tensor Potentials as a Promising Tool to Study Diffusion Processes}


I.I. Novoselov$^{1,2}$, A.V. Yanilkin$^{1,2}$, A.V. Shapeev$^3$, E.V. Podryabinkin$^3$
\vspace{5mm}

$^{1}$ Dukhov Research Institute of Automatics (VNIIA)\\
{\it 127055, Russian Federation, Moscow, Syshevskaya str. 22}
\vspace{5mm}

$^{2}$ Moscow Institute of Physics and Technology (State University) \\
{\it 141700, Russian Federation, Moscow region, Dolgoprydny, Institytski str. 9}

$^{3}$ Skolkovo Institute of Science and Technology \\
{\it 143026, Russian Federation, Moscow, Skolkovo Innovation Center, 3}

\vspace{5mm}

E-mail: novoselov@vniia.ru, novoselov92ivan@gmail.com (I.I. Novoselov)

\end{center}

\section*{Abstract}


A recently proposed class of machine-learning interatomic potentials --- Moment tensor potentials (MTPs) --- is investigated in this work. MTPs are able to actively select configurations and parametrize the potential on-the-fly. It is shown that MTPs accurately reproduce energies, forces and stresses calculated ab initio. As a more comprehensive test, MTPs are employed to calculate vacancy diffusion rates in Al, Mo and Si. We demonstrate that the results are in a good agreement with ab initio data for the materials considered.

Keywords: {\it Moment tensor potentials; active learning; diffusion; aluminum, molybdenum, silicon.}

\newpage

\section{Introduction}
Molecular dynamics (MD) has proved itself as a useful and often irreplaceable tool for various areas of research, such as materials science, chemistry and biology. The core of this method is the employed model of interatomic interactions, it determines a delicate balance between computational cost of the simulation and fidelity of the results.

One of the most accurate description of interatomic interactions is provided by quantum-mechanical models, such as density functional theory (DFT). However, applicability of DFT is limited to modelling of several hundreds of atoms at sub-nanosecond time intervals.

Atomistic simulations at larger time and space scales are often performed with  semi-empirical interatomic potentials. Such a potential has a pre-defined functional form and a number of adjustable parameters. The parameters are fitted to DFT  (and sometimes experimental) data in order to describe a particular material. This technique is computationally efficient, but often yields only qualitative results.

There is a number of approaches that aim to develop models with intermediate characteristics: less computationally intensive than DFT, but also more accurate than semi-empirical potentials. One of the approaches is machine-learning interatomic potentials (MLIPs). On one hand, they inherit some general approximations typical for semi-empirical potentials, e.g. they are local and energy of the system is represented as a sum of atomic contributions. This enables computational efficiency of the model. On the other hand, MLIPs have very flexible functional form that allows one to achieve more accurate description of interatomic interactions.


High accuracy along with affordable computational cost makes MLIPs a promising tool for materials modeling. Moreover, some MLIPs are able to actively learn on-the-fly, in other words, the potential is automatically re-fitted in order to account for new configurations when substantial extrapolation is detected. This feature is very attractive for investigation of rare events, e.g. diffusion at fairly low temperatures. In this case, the system spends most of the time in the same area of the phase space. Therefore, instead of performing a lot of DFT calculations of similar states, it is more efficient to parametrize the local potential landscape with a MLIP. Moreover, when the transition event eventually occurs, a MLIP is expected to detect extrapolation and invoke a DFT code.

Learning on-the-fly is supported by the moment tensor potentials (MTPs) --- a recently proposed class of MLIPs \cite{Shapeev_mtp}. In this work, we assess applicability of MTPs to simulation of rare events on the example of vacancy-driven diffusion in aluminum, molybdenum and silicon. Note that these materials have significantly different properties (e.g. lattice symmetry and band structure). This is done intentionally in order to obtain a more comprehensive picture.

The paper has the following structure. The concept of MTPs and calculation details are discussed in the Methods section. The first part of the Results and Discussion section is dedicated to the investigation of the effect of internal parameters on the quality of MTPs. Then the accuracy of MTPs is compared with that of semi-empirical potentials. Subsequently, MTPs are employed to calculate vacancy diffusion rates in Al, Mo and Si. The results are discussed and compared with the existing data in the final part of the paper.

\section{Methods}
Moment tensor potentials are briefly discussed in the first part of this section, while the second part is dedicated to the details of DFT calculations.

\subsection{Moment tensor potentials}

Like the absolute majority of interatomic potentials, MTP implies that the total interaction energy of a configuration can be represented as a sum of atomic contributions. The contribution from atom $i$  can be defined as $V(\boldsymbol{r_i})$, where $V$ is the interatomic potential and $\boldsymbol{ r_i}=(r_{i, 1}, ..., r_{i,n})$ is a collection of vectors pointing from the atom $i$ to its neighbors inside the potential cut-off. MTP then postulates linear representation of each of the atomic contributions $V(\boldsymbol{r_i})$:

\begin{equation}
V(\boldsymbol{r_i}) = \sum_{j=1}^m \theta_j B_j(\boldsymbol{r_i}),
\label{mtp_repres}
\end{equation}
where $\theta_j$ are adjustable parameters, $B_j$ are pre-defined basis functions and $m$ is the number of functions in the basis. Functional form of $B_j$ and other details on how the basis is constructed can be found in \cite{Shapeev_mtp, Shapeev_active}.

The total energy of the system is given by the sum of $V(\boldsymbol{r_i})$ over all atoms:

\begin{equation}
E(x) = \sum_{i=1}^N \sum_{j=1}^m \theta_j B_j(\boldsymbol{r_i}),
\label{mtp_energy}
\end{equation}
where $N$ is the number of atoms in the configuration $x$. The force acting on $j$-th atom $f_j(x)$ is determined as a derivative of $E(x)$ with respect to the atom position $x_j$:
\begin{equation}
f_j(x) = - \nabla_{x_j} E(x).
\label{mtp_force}
\end{equation}
Likewise, the virial stresses can be found as derivatives of $E(x)$ with respect to the lattice vectors $L$:
\begin{equation}
\sigma(x) = \frac{1}{|{\rm det}(L)|} (\nabla_{L} E(x)) L^\top.
\label{mtp_stress}
\end{equation}


From Eq.\ref{mtp_repres}--\ref{mtp_stress} it follows that the energy, forces and stresses for a given configuration are determined by the set of basis functions, and the values of the adjustable parameters $\theta_j$. These parameters are found through the fitting to the results of DFT calculations. Imagine that the calculations were performed for a collection of configurations $X_{TS}$, which we denote as the training set. 

For each $x_i$ in $X_{TS}$, we know the ``exact'' energy ($E^{DFT}(x_i)$), per-atom forces ($f^{DFT}_j(x_i)$) and the components of the stress tensor ($\sigma^{DFT}_j(x_i)$). Therefore, the MTP energy error for $x_i$ is given by:
\begin{equation}
 \Delta E(x_i) = |E(x_i) - E^{qm} (x_i) |.
\label{mtp_dE}
\end{equation}
The errors of forces and stresses are defined in a way similar to Eq.\ref{mtp_dE}. Then the values of $\theta_j$ can be found through minimization of the functional:
\begin{equation}
\sum_{x_i \in X_{TS}} \left[ C_E^2 \Delta E(x_i) ^2 + C_f^2 \sum_{j=1}^{N_i} \Delta f_j(x_i)^2 + C_{s}^2 \Delta \sigma(x_i)^2 \right].
\label{mtp_func}
\end{equation}
Note that the energy, force and stress terms in Eq.\ref{mtp_func} are weighted by $C_E$, $C_f$ and $C_s$, respectively. The weigh factors allow one to determine the relative importance of energies, forces and stresses during the fitting routine. Further, we denote these parameters as the fitting weights.


MLIPs are also known to be sensitive to the nature of configurations included in the training set. The highly flexible functional form of MLIPs limits their transferability, i.e. the ability to extrapolate. Therefore, the optimal training set should cover the whole phase space of the system without significant ``gaps'' to avoid extrapolation. The process of fitting the potential to an optimized training set is called active learning.

Active learning is readily supported by MTPs \cite{Shapeev_active}. The D-optimality criterion is employed in this case in order to decide whether to include a configuration into the training set. Note that each configuration can be represented as a point in the phase space, and a set of configurations forms a simplex. Therefore, the criterion has a transparent physical interpretation: a configuration is included into the training set, if it increases the simplex volume. Importantly, the decision is made based only on atomic coordinates. This feature, along with the linear form of the potential (Eq.\ref{mtp_repres}), allows MTP to effectively learn on-the-fly.


MTP is able to select configurations for the training set based on several features: neighbors, energies, forces and stresses. The strategies could be combined together using the selection weights, like in the fitting process. Besides the weights, the selection is controlled by a threshold value. The selection threshold determines the minimal increase in the volume of the phase-space simplex required to add a configuration to the training set. Hence a decrease of the threshold leads to an increase in the number of selected configurations.

To sum up, the parametrization of MTPs includes the following steps:

\begin{itemize}
\item First of all, a training set should be selected from a given database of configurations (or from MD trajectory on-the-fly). The selection process is governed by a threshold and four selection weights: neighbor, energy, force and stress.

\item Then the potential is fitted to the training set. The fitting process is controlled by three fitting weights: energy, force and stress.
\end{itemize}

The effect of the fitting weights and selection parameters will be discussed in details in Section \ref{results_section}.
\subsection{DFT calculations}

Like many other interatomic potentials, MTPs are fitted to the results of quantum-mechanical calculations in the framework of density functional theory (DFT). DFT calculations were performed using the VASP code \cite{VASP_cite_1}. We employed projector augmented wave potentials \cite{PAW_definition, PAW_potentials} generated with {\it Perdew, Burke} and {\it Ernzerhof} generalized gradient approximation \cite{GGA_PBE_1996}. 3, 4 and 6 electrons were treated as valence (out-of-core) states for Al, Si and Mo, respectively.

The plain wave basis cutoff ($E_{cut}$) and the number of k-points were determined based on convergence tests for energies and forces. The tests indicated that the $E_{cut}$ of 300 eV  is sufficient for Al, and 400 eV for Mo and Si. In addition, the tests suggested that $3 \times 3 \times 3$ k-mesh is required for all of the materials.

Note that Al, Si and Mo have different lattice symmetries: FCC, diamond and BCC, respectively. Atoms were placed in computational cells in order to form appropriate lattice, a vacancy was also created in each of the cells. As a result, they contained 107, 63 and 53 atoms for Al, Si and Mo, respectively. Temperature-dependent lattice parameters were adopted from the X-ray experiments \cite{Al_expans, Si_expans, Mo_expans}.

Databases of DFT configurations were required in order to select a set of optimal internal parameters of MTP.  Equilibrium quantum MD trajectories were calculated for this purpose. Each of them contains $2 \cdot 10^4$ frames generated with a timestep of 1 fs. Temperature during the MD runs was maintained at $0.9 \; \mbox{T}_{melt}$ by means of the Nos\'{e} thermostat. More specifically, the dynamics of Al, Si and Mo was simulated at 850, 1650 and 2600 K, respectively.

\subsection{Calculation of diffusion coefficients}

MTP was employed to calculate the vacancy diffusion coefficients in Al, Si and Mo. If not stated otherwise, this was done in the following way. 

A separate MTP is parametrized for every temperature investigated. At a particular temperature, MTP potential is fitted via the active learning on-the-fly algorithm, the fitting weights are 100, 10 and 1 for energy, force and stress, respectively. The selection of configurations is based on neighbors and forces, the corresponding weights are 1 and 10, accordingly. The length of the learning trajectory is $10^6$ steps, the timestep, as usual, is $1 \; fs$.

Consequently, the learning is switched off and the obtained potential is employed to conduct four independent MD runs, $5 \cdot 10^6$ steps each. The vacancy diffusion coefficients are then calculated for each of the runs in accordance with the Einstein-Smoluchowski relation. Several runs are performed in order to obtain better statistics. The number of runs can was increased up to 16, if significant scatter of diffusivities was observed, which is usually the case for low temperatures.

Vacancy contribution to self-diffusion is then calculated in a standard way:
\begin{equation}
D_{self} = f D_v c_v(T),
\label{eq_selfdiff}
\end{equation}
where $f$ is the correlation factor which equals to $0.7815$, $0.7215$ and $0.5$ for Al, Mo and Si, respectively \cite{Leclaire1955, Blochl1993};  $D_v$ is the vacancy diffusion coefficient; $c_v(T)$ is the temperature-dependent vacancy concentration. Vacancy concentrations that account for anharmonic effects were adopted from \cite{Glensk2014} (Al) and \cite{Mattsson2009} (Mo). Considering Si, only an Arrhenius interpolation of $c_v(T)$ is available in the literature, hence we used the parameters determined in \cite{Voronkov2006}.

\section{Results and Discussion}
\label{results_section}
MTP is a relatively new class of machine-learning interatomic potentials. Therefore, it is interesting to test its ability to reproduce DFT data at first. Below, we consider energy, force and stress errors in order to obtain a comprehensive picture.

As discussed in the Methods section, MTP has several internal parameters. Three of them (fitting weights) control the parametrization process, and five other options (threshold and four weights) govern the selection of the training set from a database of configurations. Let us consider the fitting weights first. 

\subsection{Effect of MTP fitting weights}

\begin{figure}[h!]
\begin{flushright}
\begin{overpic}[width=0.49\textwidth]{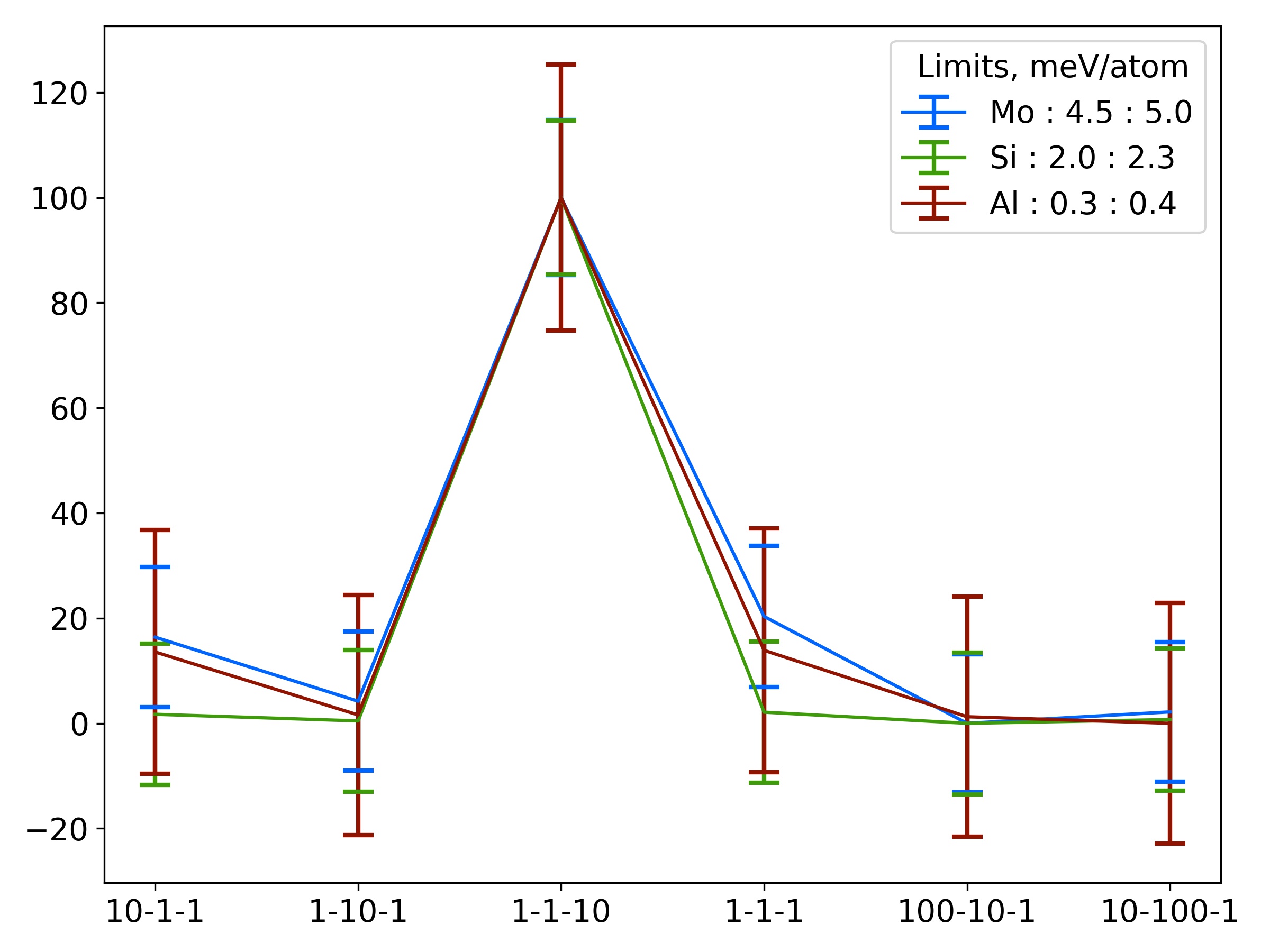}
\put(50, 77){\makebox(0,0){\bf $\Delta$ Mean abs. error}}
\put(-4, 30){\rotatebox{90}{\small Energy}}
\end{overpic}
\begin{overpic}[width=0.49\textwidth]{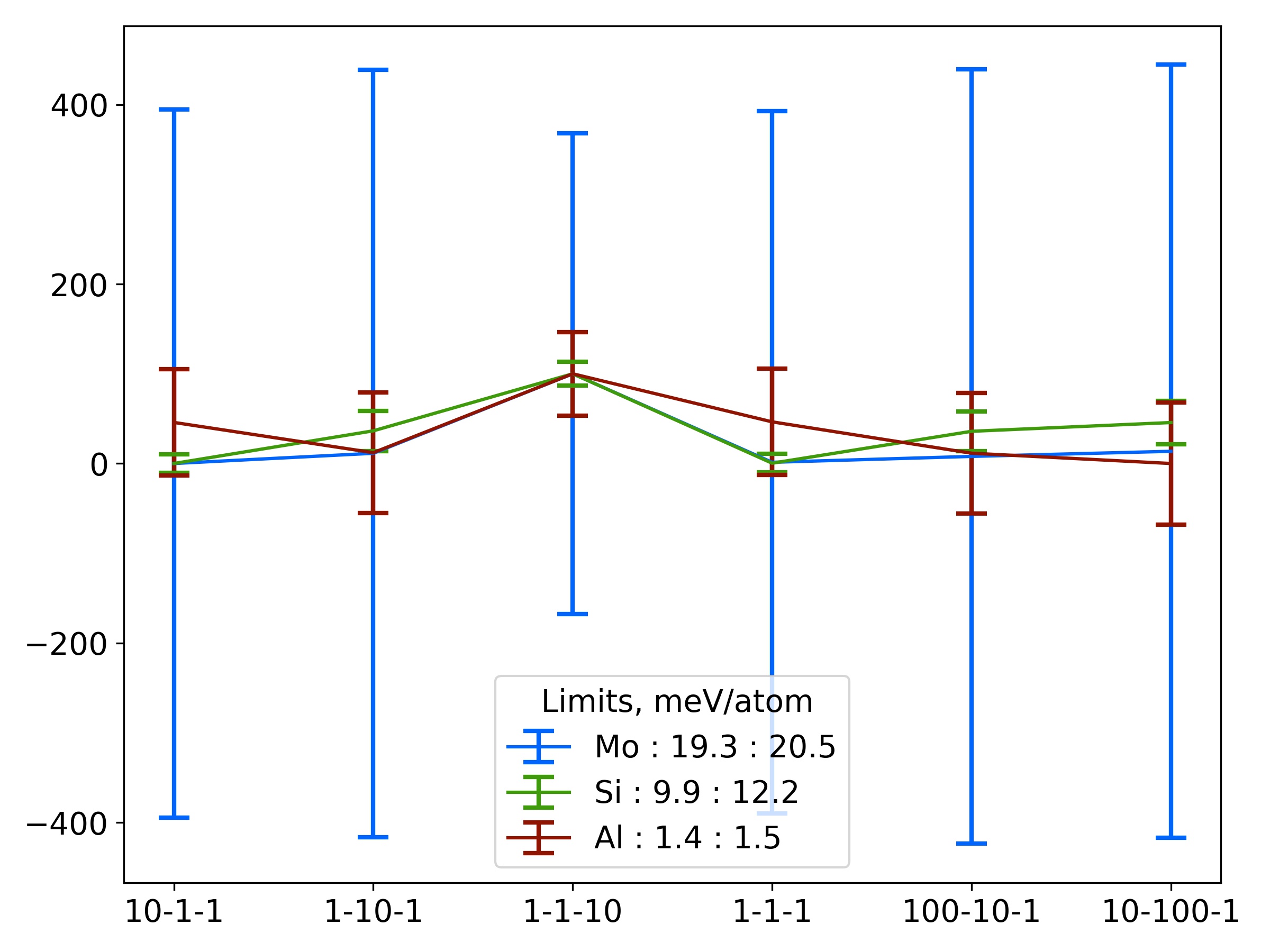}
\put(50, 77){\makebox(0,0){\bf $\Delta$ Max abs. error}}
\end{overpic}
\begin{overpic}[width=0.49\textwidth]{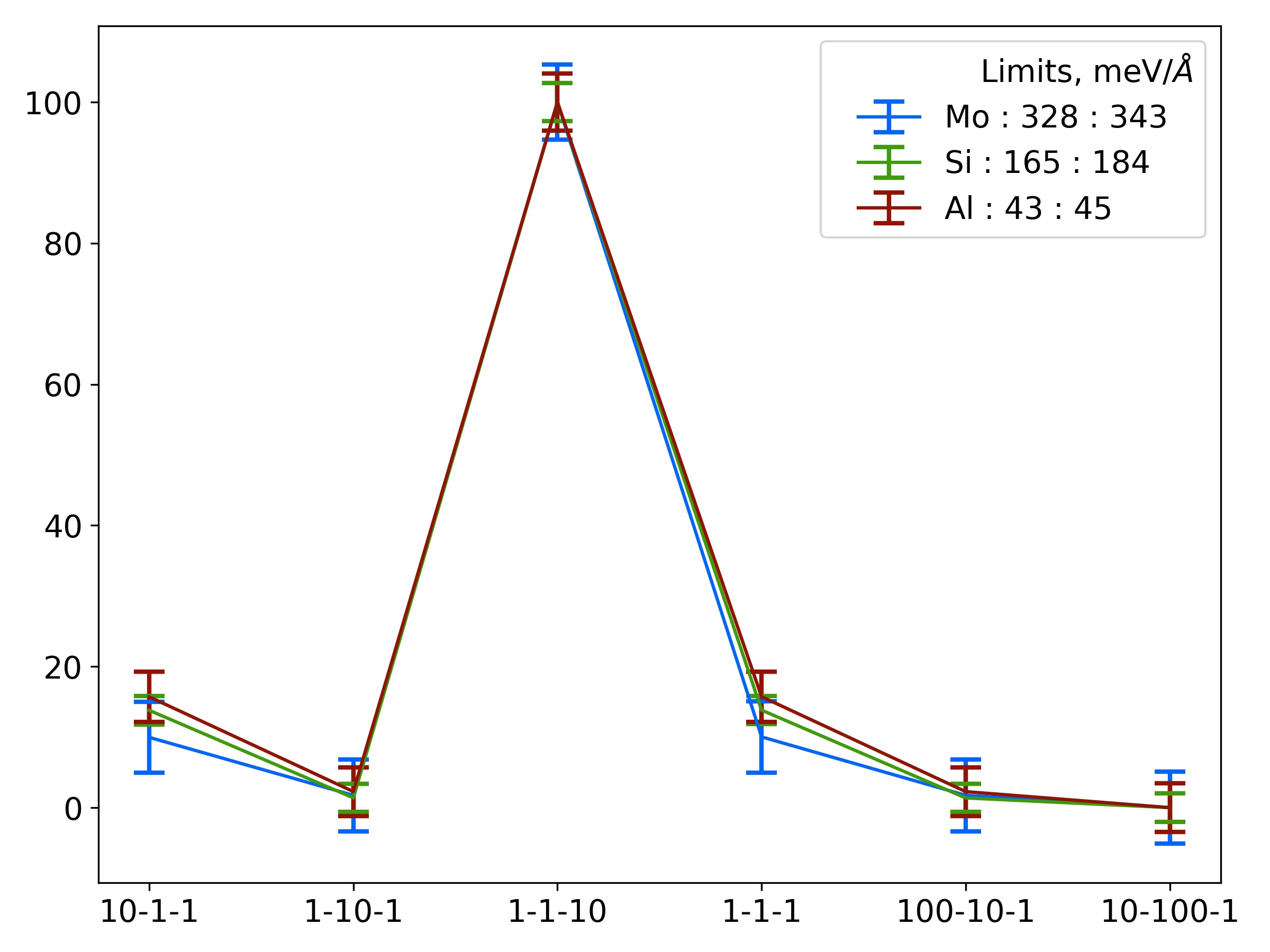}
\put(-4, 30){\rotatebox{90}{\small Force}}
\end{overpic}
\begin{overpic}[width=0.49\textwidth]{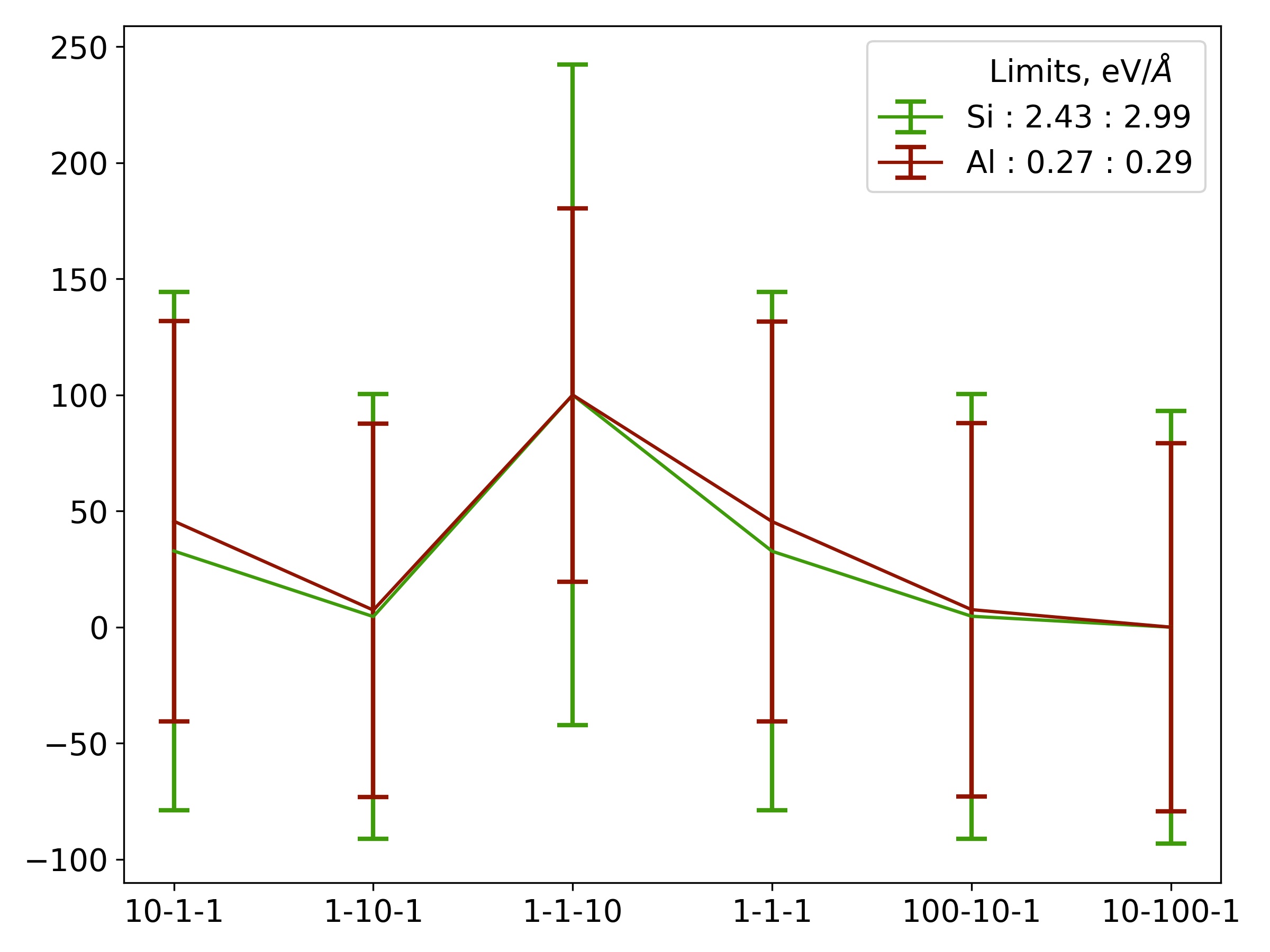}
\end{overpic}
\begin{overpic}[width=0.49\textwidth]{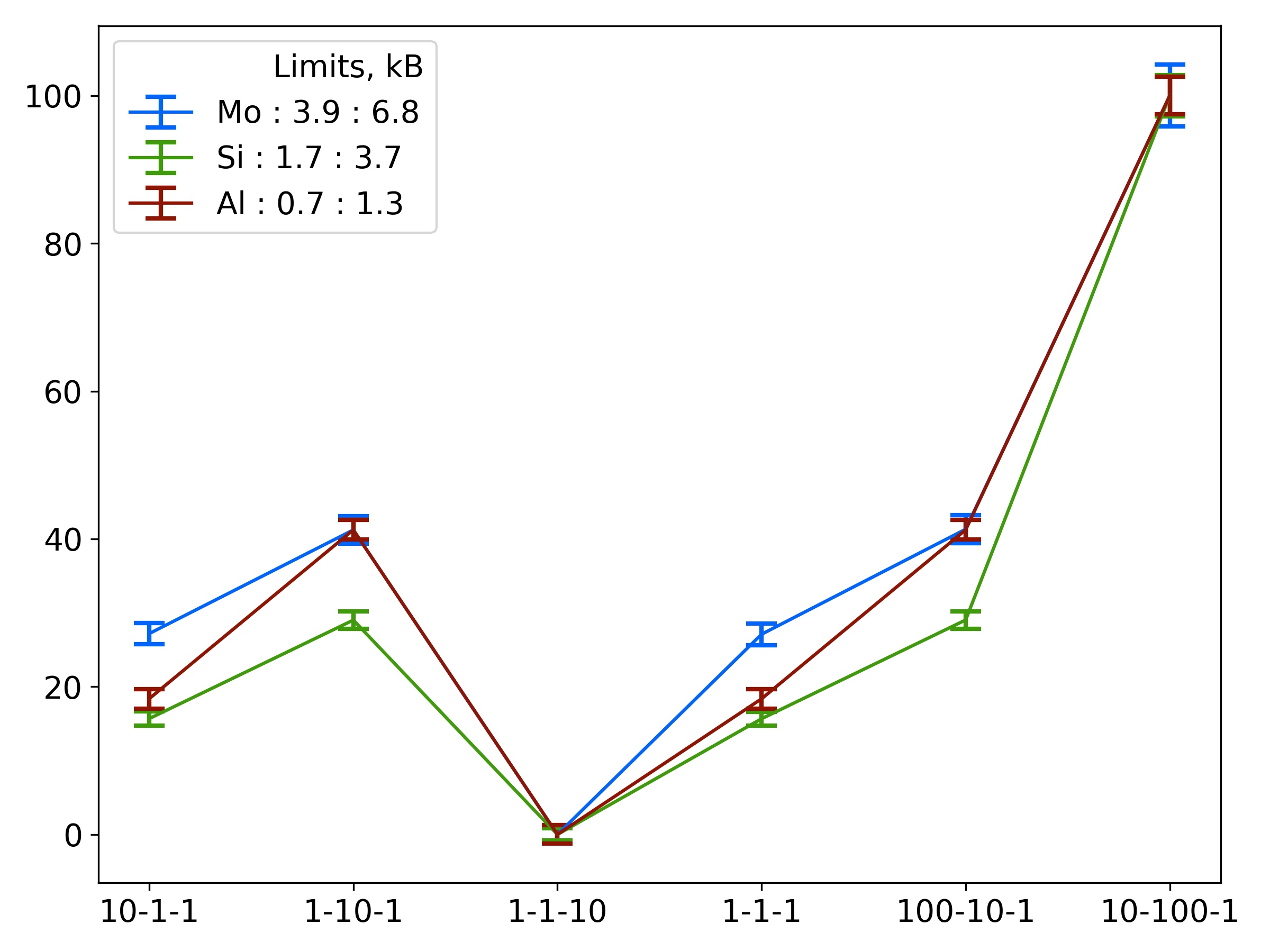}
\put(-4, 30){\rotatebox{90}{\small Stress}}
\put(50, -2){\makebox(0,0){\small Fit weights}}
\end{overpic}
\begin{overpic}[width=0.49\textwidth]{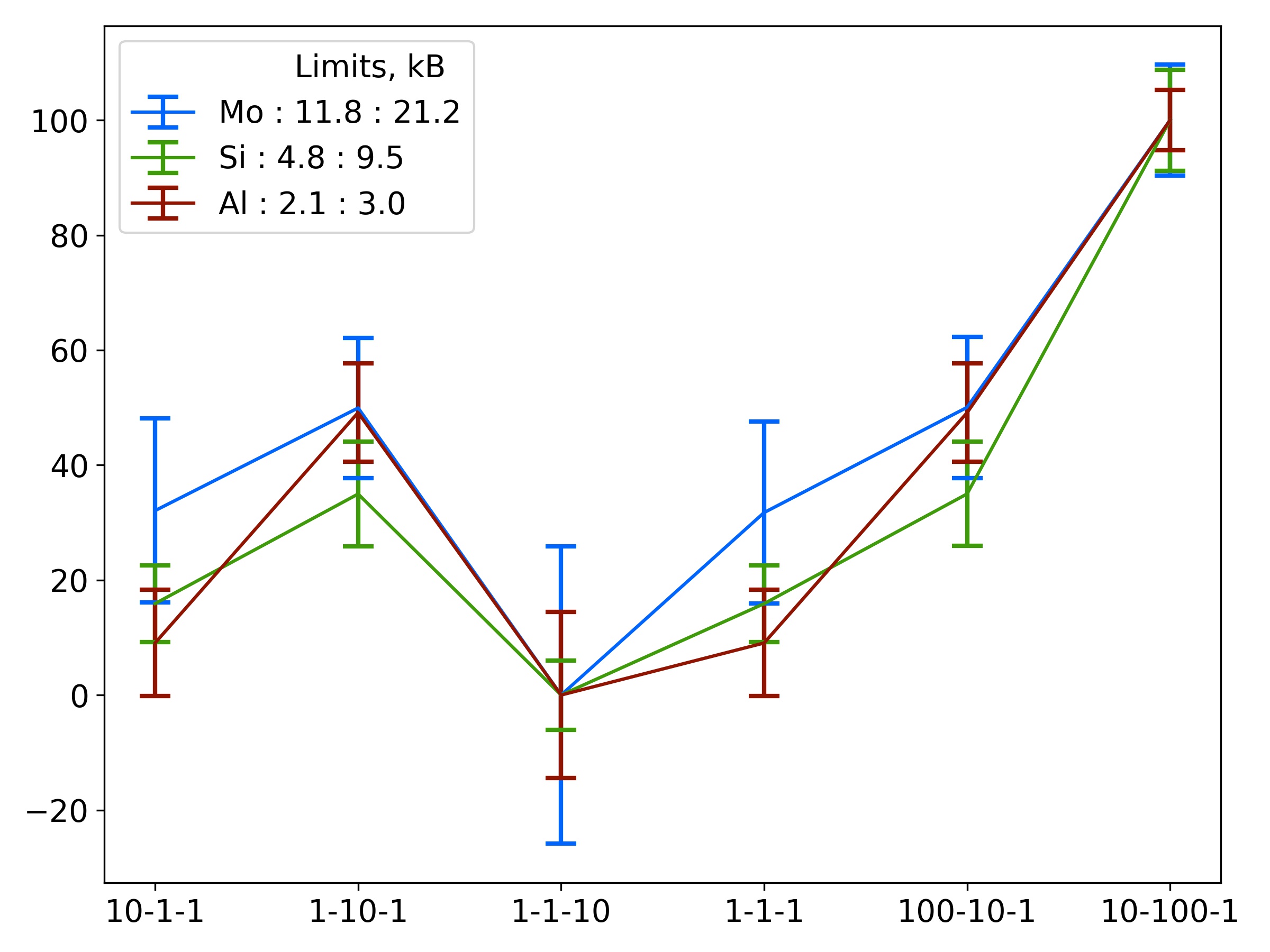}
\put(50, -2){\makebox(0,0){\small Fit weights}}
\end{overpic}
\end{flushright}

\caption{MTP errors for energies (first row), forces (second row) and stresses (third row) for various fit weights. Left column: relative mean absolute errors, right column: relative maximum absolute errors. The numbers in legends indicate min and max values for all of the weights considered. Y-axis units are chosen in order to remap errors to the range from 0 to 100. Max abs. error for molybdenum forces is not shown due to large statistical uncertainty (over 10 times), corresponding average value remains between $7.8$ and $8.4 \; eV/\mbox{\AA}$.}
\label{fig:EFS_passive}
\end{figure}

A natural way to assess the effect of fitting weights is to train MTP potentials with different weight sets and compare their accuracy afterwards. To that end, train and test datasets are required. The datasets were generated by random selection of 2000 unique configurations out of 20 ps MD trajectory (see Methods for details). 64 pairs of non-overlapping train and check datasets were generated this way in order to guarantee statistical reliability of the results. 


The data is plotted on Fig.\ref{fig:EFS_passive}, standard deviations of the values are indicated with error bars. The figure contains six plots: mean and maximum absolute errors of energies, forces and stresses. Each of the plots (except for one, see the caption) demonstrates the results for molybdenum, silicon and aluminum. Note that the accuracy of MTP strongly depends on the material investigated. Hence, the errors were remapped to the range from 0 to 100 in order to facilitate the comparison. The minimal and maximal values of the errors are given in legends.

The considered fitting weights are indicated on the X-axis on each of the plots in the following notation: ``energy weight'' - ``force weight'' - ``stress weight''. For example ``100-10-1'' means that the energy weight equals to 100, the force --- to 10 and the stress --- to 1.

It is clear from Fig.\ref{fig:EFS_passive} at first glance that the effect of fitting weights is not large. Indeed, the variations of energy and force errors are typically in the range from 10 to 30 \%. The changes of stress errors are a bit more pronounced, but still within a factor of 2. Nevertheless, even a 10 \% improvement could be important in some cases, let us therefore have a closer look at mean absolute errors. 

One can see from the left column of Fig.\ref{fig:EFS_passive} that the effect of fitting weights on mean-absolute errors is similar for all of the materials.  In particular, high value of the stress weight, as expected, decreases the stress error. However, it also makes energy and force errors grow. Compare the sets ``1-1-1'' and ``1-1-10'', or ``1-10-1'' and ``10-100-1'' for example. 

Interestingly, this tendency is not observed for mean energy errors. Note that the error is higher for the ``10-1-1'' set (at least for Mo and Al) than for the ``1-10-1''. Hence, accurate description of energy requires the force weight to be significantly higher than the stress weight.

It should be pointed out that accurate description of forces and energies has higher priority, since the potential will be employed to perform MD simulations. Consequently, ``1-10-1'' and ``100-10-1'' are good candidates for the optimal fitting weights. Note that we do not consider  ``10-100-1'' here, because it maximizes the stress error without significant gain in accuracy of energies or forces.

As can be seen from the right column of Fig.\ref{fig:EFS_passive},the maximum absolute errors demonstrate the same features as mean absolute errors. However, standard deviation of maximum errors is much larger than that of the mean errors. This effect is especially pronounced for molybdenum force errors. Their standard deviation is approximately  $8.4 \; eV/\mbox{\AA}$, while the mean values vary from $7.8$ to $8.4 \; eV/\mbox{\AA}$. 

Therefore, looking at the error bars on Fig.\ref{fig:EFS_passive}, one can conclude that fitting weights have no appreciable effect on maximum absolute errors of forces and energies. Regarding stresses, the errors behave similarly to the mean absolute errors.

In summary, we analyzed the effect of fitting weights on the accuracy of MTP potentials. The obtained results suggest that the weight sets of  ``1-10-1'' and ``100-10-1'' yield the most accurate MTP. In order to choose between the sets, it should be noted that each DFT calculation yields only one energy, but three force components per every atom in the system. Therefore, it seems better to keep the energy weight larger than the force weight to avoid force overfitting. That is why we will employ the ``100-10-1'' set in future, albeit no signs of overfitting were observed in the tests.

\subsection{Effect of MTP selection weights}

Now, when the optimal values of fitting weights are established, it is appropriate to consider the effect of a selection strategy on the quality of MTP. By quality we mean not only accuracy, but also predictive power of the potential. Therefore, ten-fold cross-validation was employed instead of random sampling in this case. In other words, the whole trajectory ($2 \cdot 10^4$ steps) was split into 10 even parts. Then nine of the parts were used for selection and training, and the remaining part was employed for estimation of errors (the test set). Note that there are 10 ways of selecting the test set, therefore the whole select-train-check loop was performed 10 times. These data is then used to compute the average and the standard deviation of the test error across the iterations.

Here it is worth to remind that the selection process in MTP is controlled by five parameters, namely a threshold and four weights: neighbor, energy, force and stress.

Note that the number of selected configurations depends not only on the threshold value, but also on a selection strategy. Consider a threshold of 2.0 for example, then around 720 configurations  will be selected for the energy-based strategy, and only about 65 for the neighbor-based selection. The selection strategies will therefore be compared for similar numbers of selected configurations rather than for similar threshold values. This comparison is very practical for the on-the-fly learning scenario, since the amount of computational resources required is mainly determined by the number of selected configurations in this case.

\begin{figure}[t!]
\begin{flushright}
\begin{overpic}[width=0.49\textwidth]{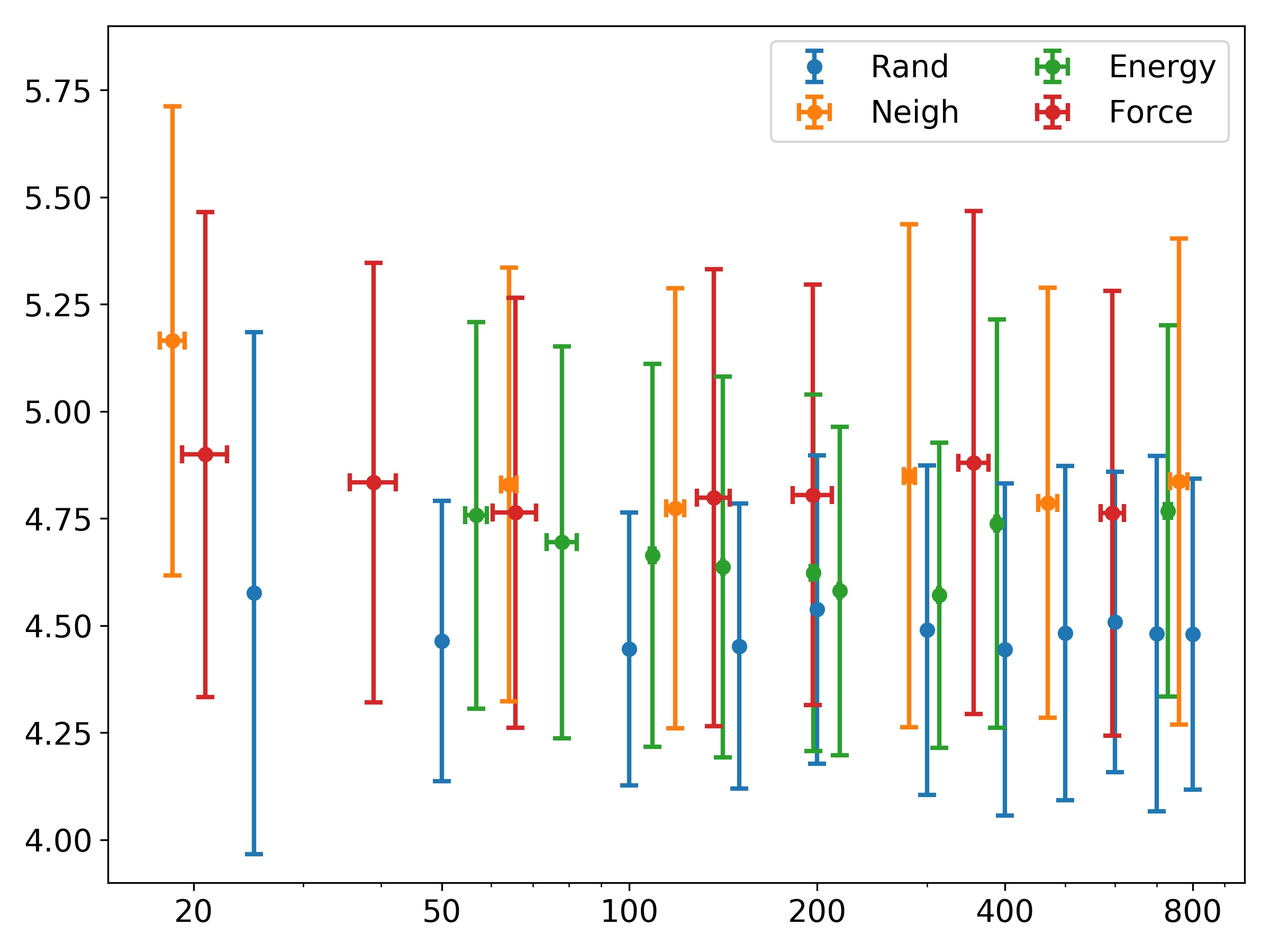}
\put(50, 77){\makebox(0,0){\bf Mean abs. error}}
\put(-6, 20){\small  \rotatebox{90}{Energy, meV/atom}}
\end{overpic}
\begin{overpic}[width=0.49\textwidth]{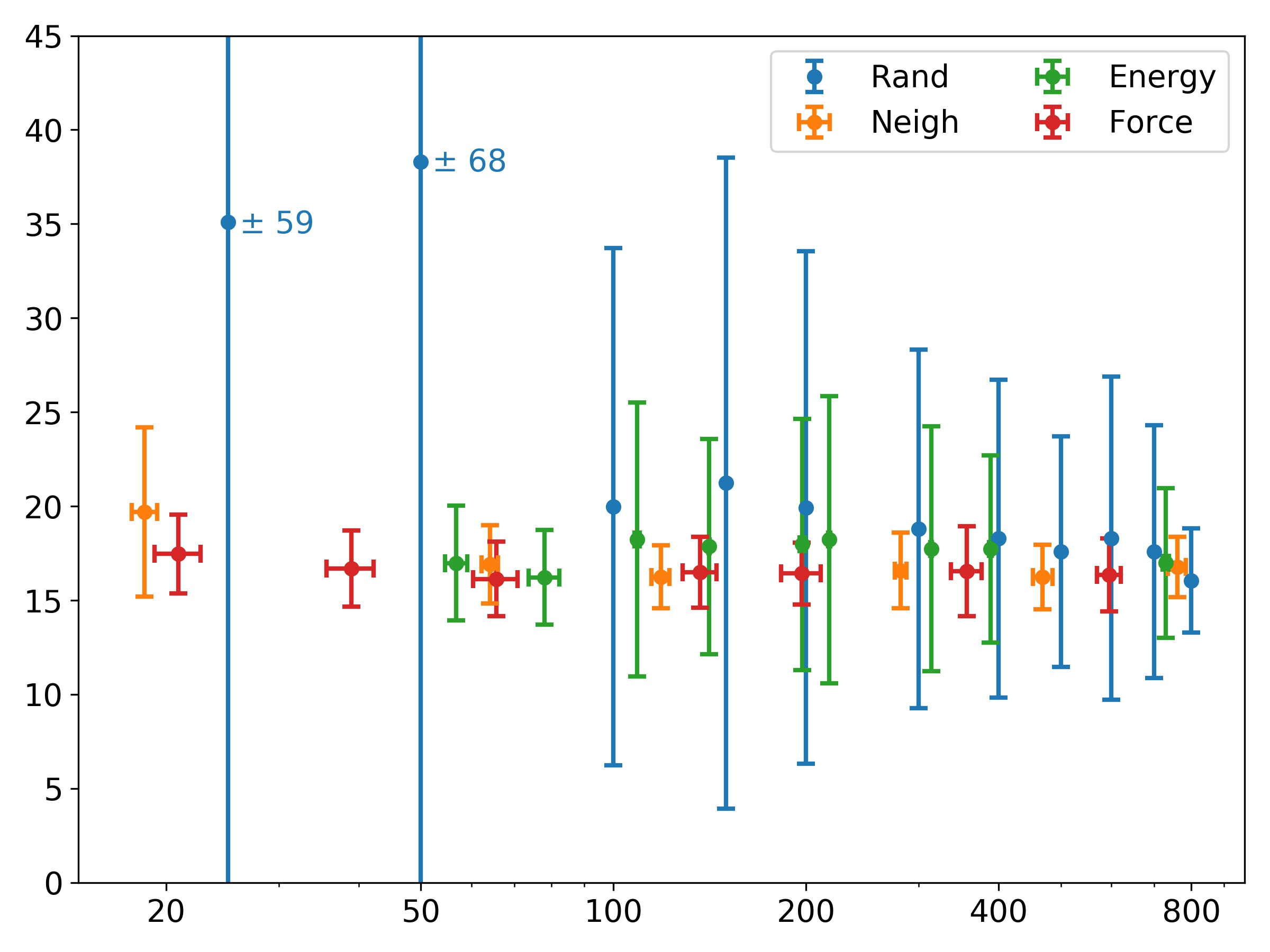}
\put(50, 77){\makebox(0,0){\bf Max abs. error}}
\end{overpic}
\begin{overpic}[width=0.49\textwidth]{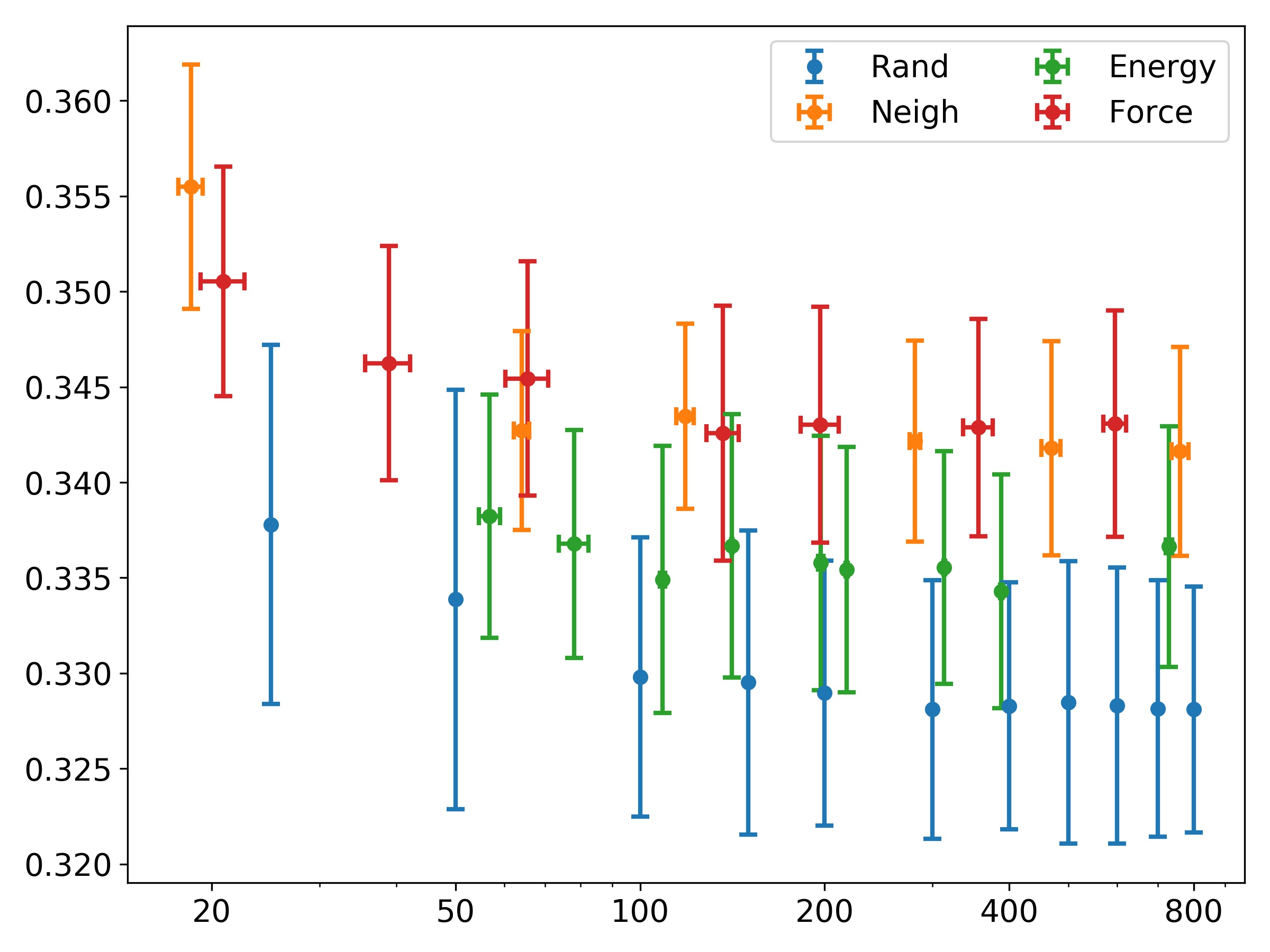}
\put(-6, 25){\small \rotatebox{90}{Force, eV/\AA}}
\end{overpic}
\begin{overpic}[width=0.49\textwidth]{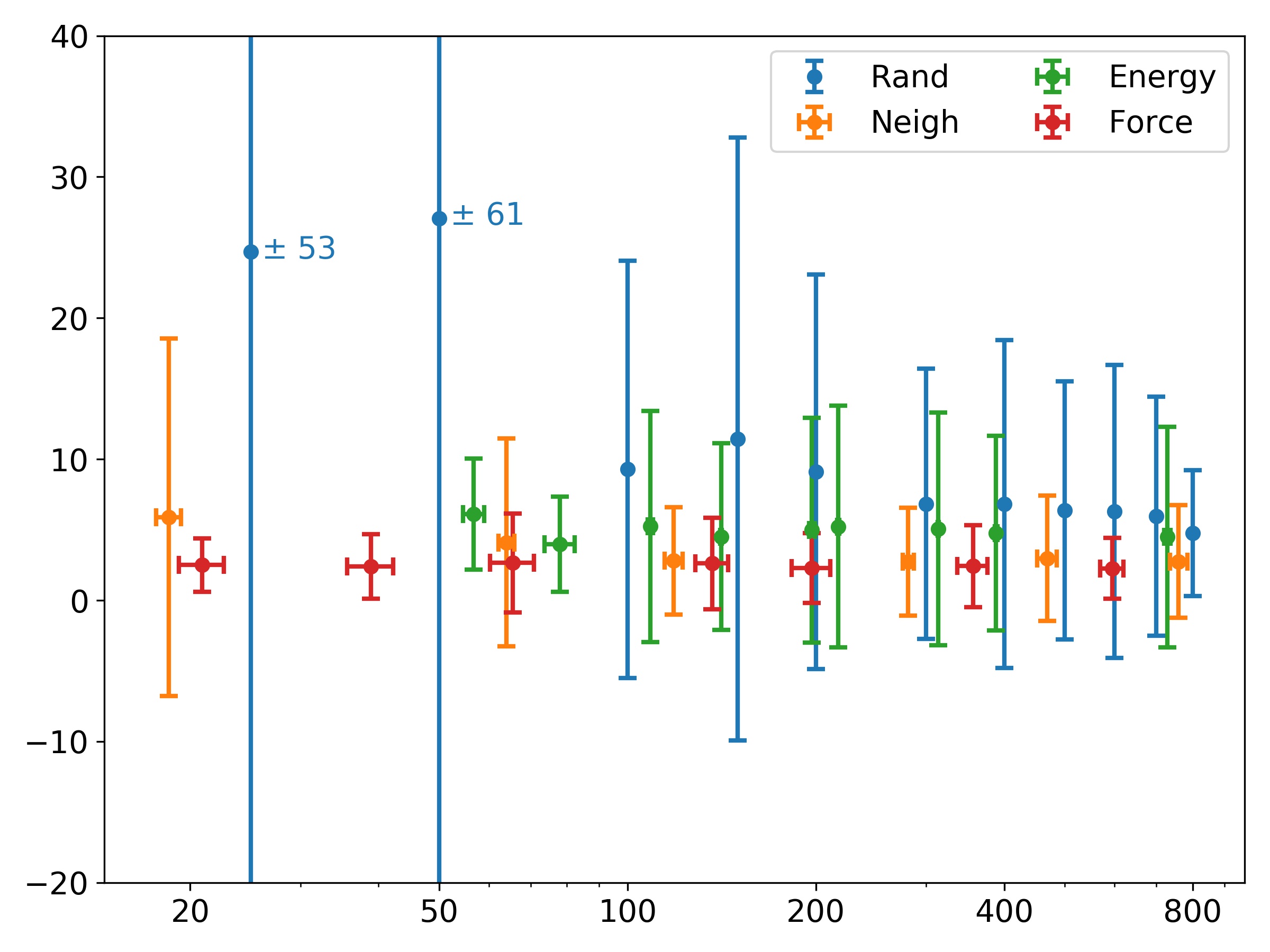}
\end{overpic}
\begin{overpic}[width=0.49\textwidth]{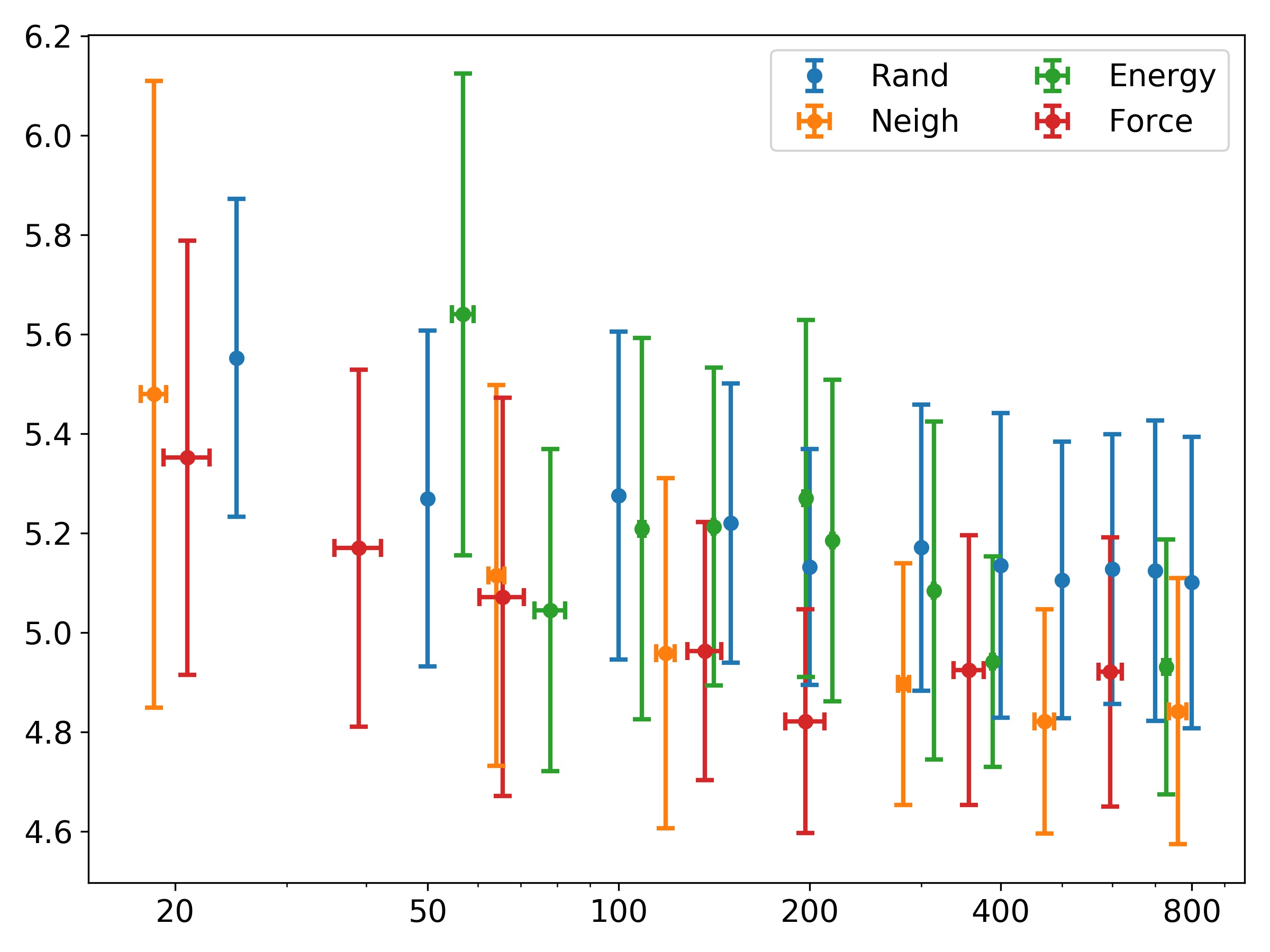}
\put(-6, 30){\rotatebox{90}{\small Stress, kB}}
\put(50, -3){\makebox(0,0){\small Num. selected}}
\end{overpic}
\begin{overpic}[width=0.49\textwidth]{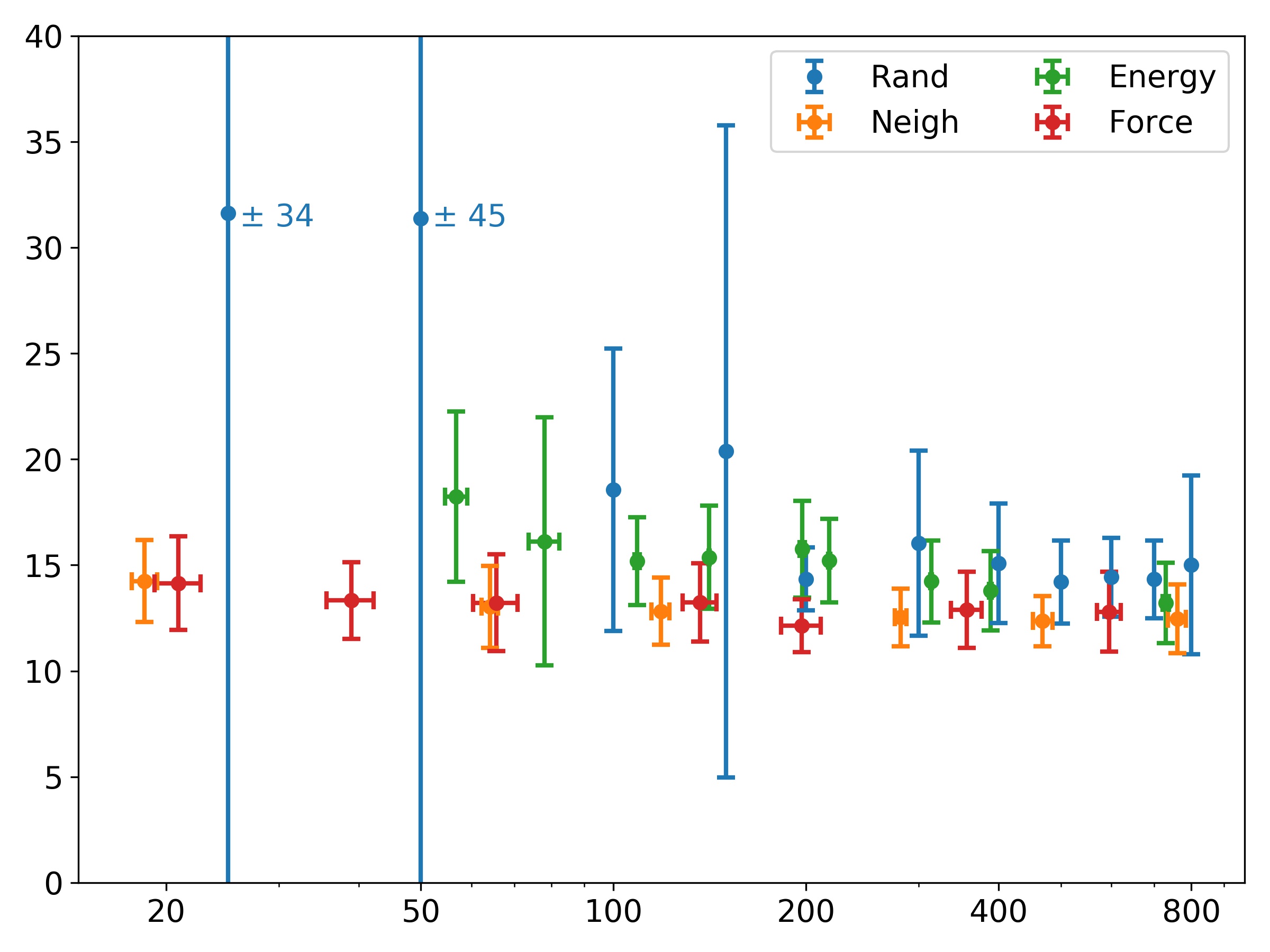}
\put(50, -3){\makebox(0,0){\small Num. selected}}
\end{overpic}
\end{flushright}

\caption{MTP errors for energies (first row), forces (second row) and stresses (third row) in molybdenum for various fit weights. Left column: mean absolute errors, right: maximum absolute errors. Note that the size of the error bars exceeds the plot range for several points, corresponding standard deviations are indicated by numeric labels in this case.}
\label{fig:EFS_active}
\end{figure}

Energy, force and stress errors obtained with different selection strategies are depicted on Fig. \ref{fig:EFS_active}. Mean and maximum absolute errors are considered, like in the case of study of the fitting weights effect. One can see from the figure that the mean energy errors are slightly lower for the random selection than for the other strategies. However, this effect is within one sigma for all of the considered numbers of selected configurations.

The difference of maximum energy errors is also within the standard deviation for various selection strategies. Nevertheless, it can be clearly seen from Fig.\ref{fig:EFS_active} that selection by neighbors or by forces allows one to significantly reduce the deviations of maximum errors. The effect is more pronounced at low numbers of selected configurations.

The same tendencies are observed for force errors on Fig.\ref{fig:EFS_active}. In particular, the neighbor and the force selection strategies demonstrate higher mean absolute errors with respect to random selection. At the same time, both of the strategies significantly reduce the standard deviation of maximum absolute errors (also with respect to random selection), especially when the numbers of selected configurations is low. Regarding the stress errors, the neighbor and the force selection strategies yield lower mean values of the mean error and, at the same time, lower standard deviations of the maximum error.

Another important feature depicted on Fig.\ref{fig:EFS_active} is that the mean force and stress errors decrease, while the number of selected configurations increases from 20 to 100--200, and then levels out. Consequently, around 200 configurations should be selected in order to optimally parametrize an MTP. These values correspond to selection thresholds in the range of 1.5--1.2 for the neighbor and the force selection strategies. However, we will employ the threshold of 1.1 in what follows in order to guarantee the highest accuracy. 

To sum up, it is demonstrated on Fig.\ref{fig:EFS_active} that the neighbor and force selection strategies allow us to minimize the maximum absolute errors (of energies, forces and stresses) and their standard deviations at the price of slight increase in mean absolute errors. The effect of the energy-based selection lies somewhere between the mentioned strategies and random selection. Therefore, the following selection weights will be employed below: neighbor 1; energy 0; force 10; stress 0.


\subsection{Comparison of MTP with semi-empirical potentials}

\label{compare}


Imagine one needs an interatomic potential in order to simulate a certain phenomenon at atomic scale. Of course one may develop a task-specific semi-empirical potential, but this is a challenging and time consuming problem itself. It is therefore more common to employ existing interatomic potential.

An alternative approach is to fit an MTP via an active learning on-the-fly algorithm. This concept is also very practical and does not require large amounts of time or computational resources.

It is now interesting to compare the accuracy of these two approaches. The comparison is made for molybdenum, since MTP provides the least accurate description of DFT data in this case. Several semi-empirical interatomic potentials for Mo were selected  \cite{Smirnova_EAM, Starikov_ADP, Park_MEAM}. The potentials have different functional forms: EAM \cite{Smirnova_EAM}, ADP \cite{Starikov_ADP} and MEAM \cite{Park_MEAM}. Note that all the potentials were parametrized based solely on DFT calculations.

\begin{figure}[h!]
\begin{flushleft}
\begin{overpic}[width=0.49\textwidth]{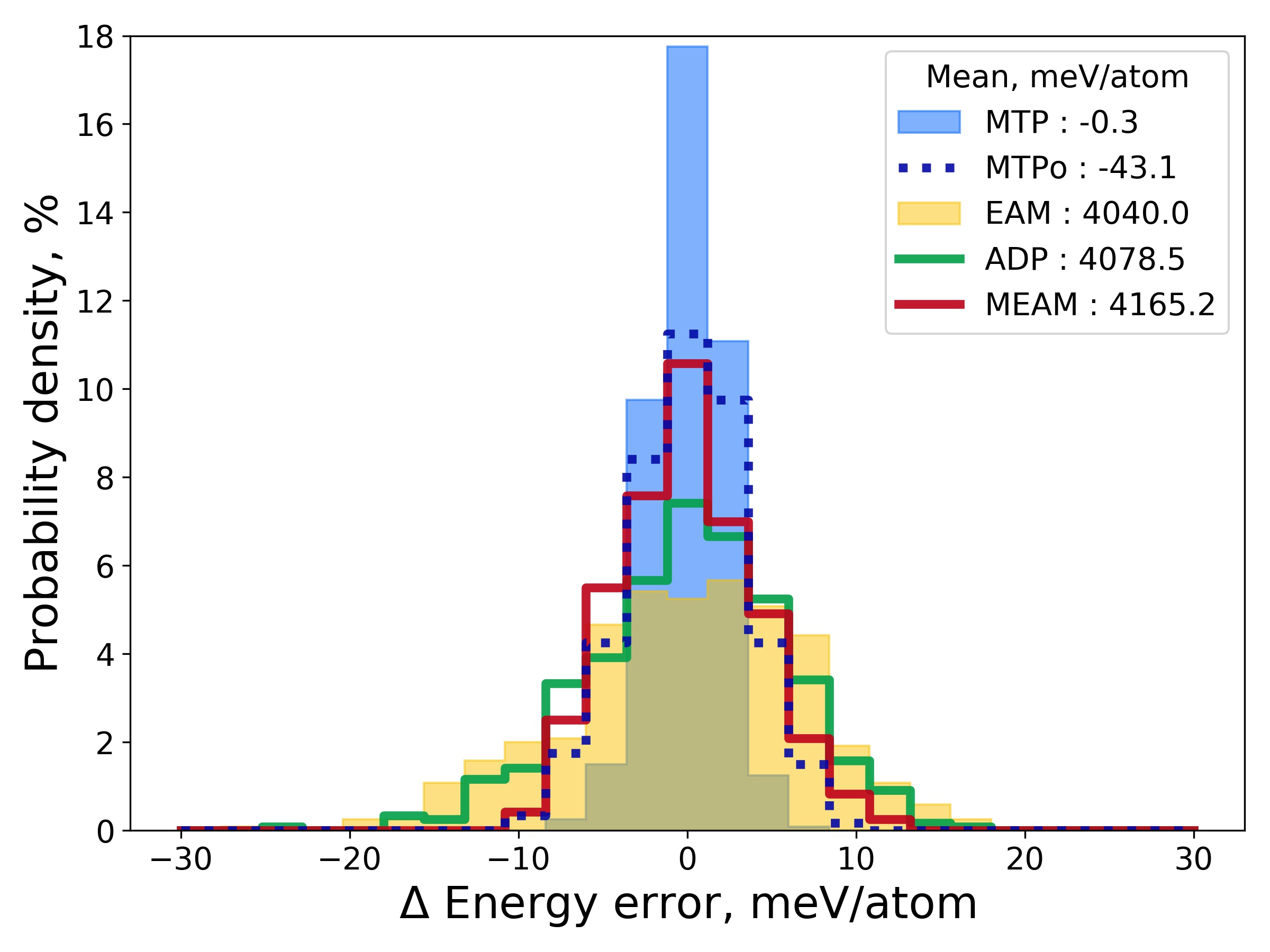}
\put(50, 77){\makebox(0,0){\bf T = 1450 K}}
\end{overpic}
\begin{overpic}[width=0.49\textwidth]{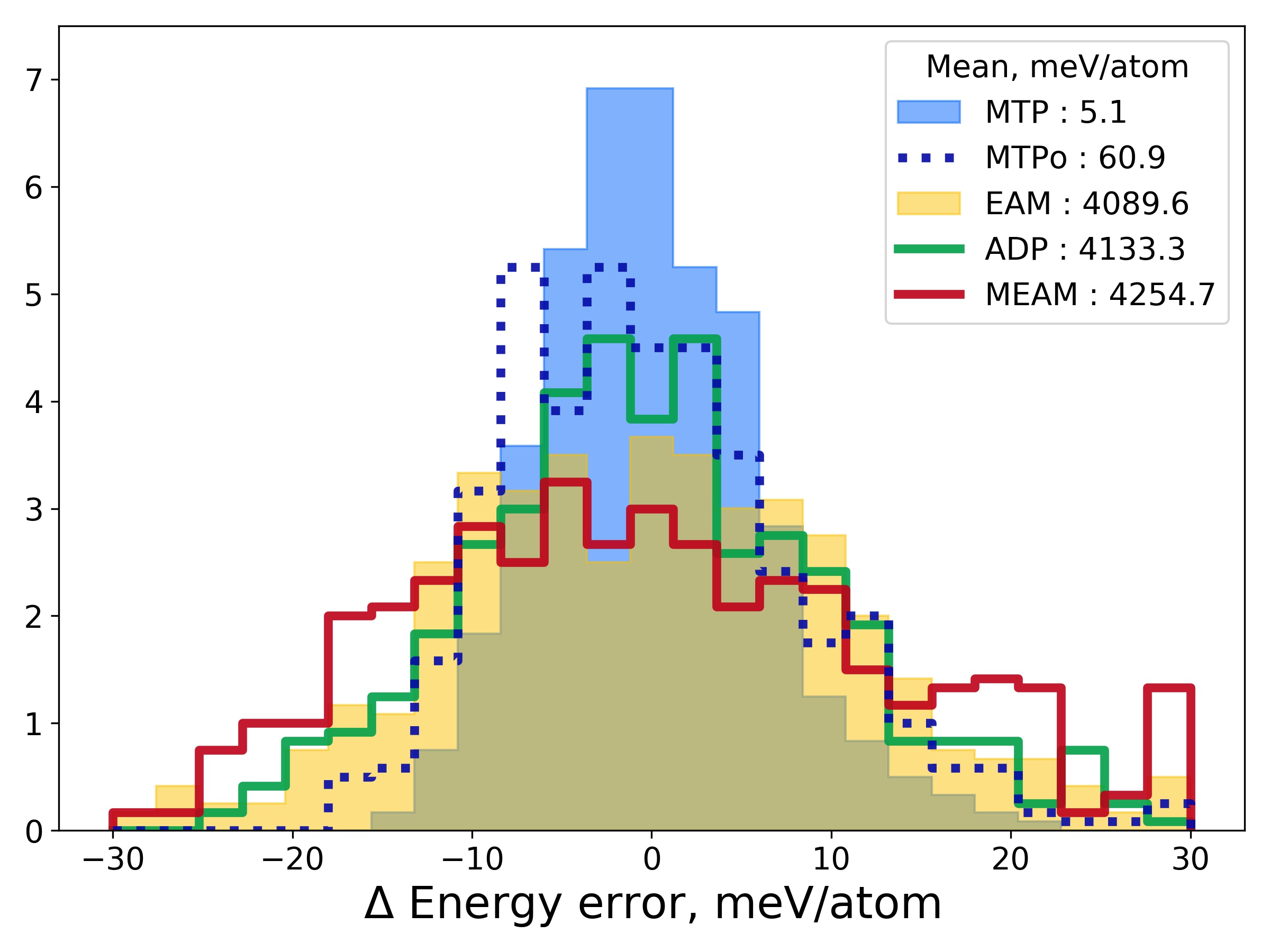}
\put(50, 77){\makebox(0,0){\bf T = 2600 K}}
\put(100, 50){\rotatebox{-90}{\small Energy}}
\end{overpic}
\begin{overpic}[width=0.49\textwidth]{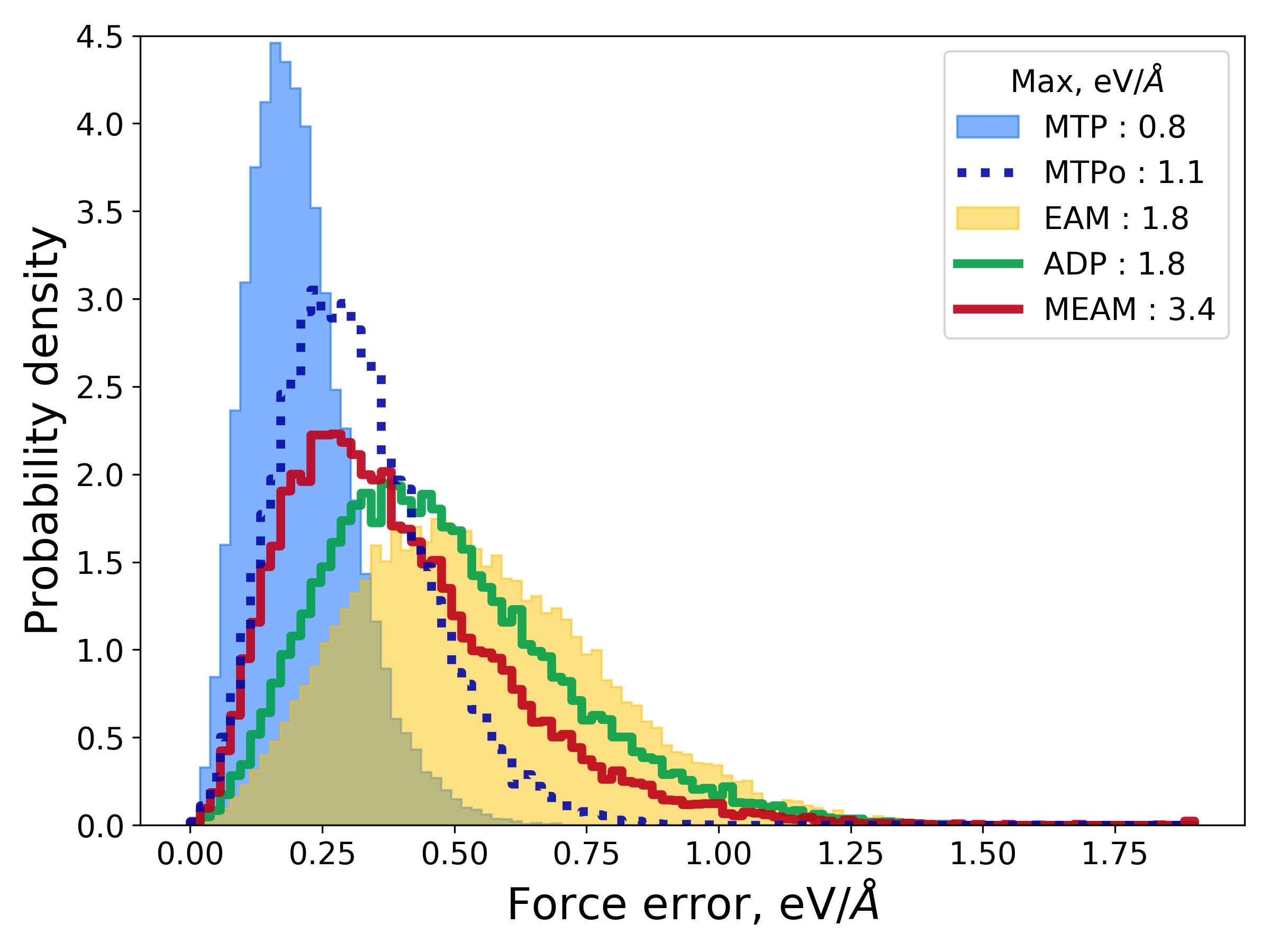}
\end{overpic}
\begin{overpic}[width=0.49\textwidth]{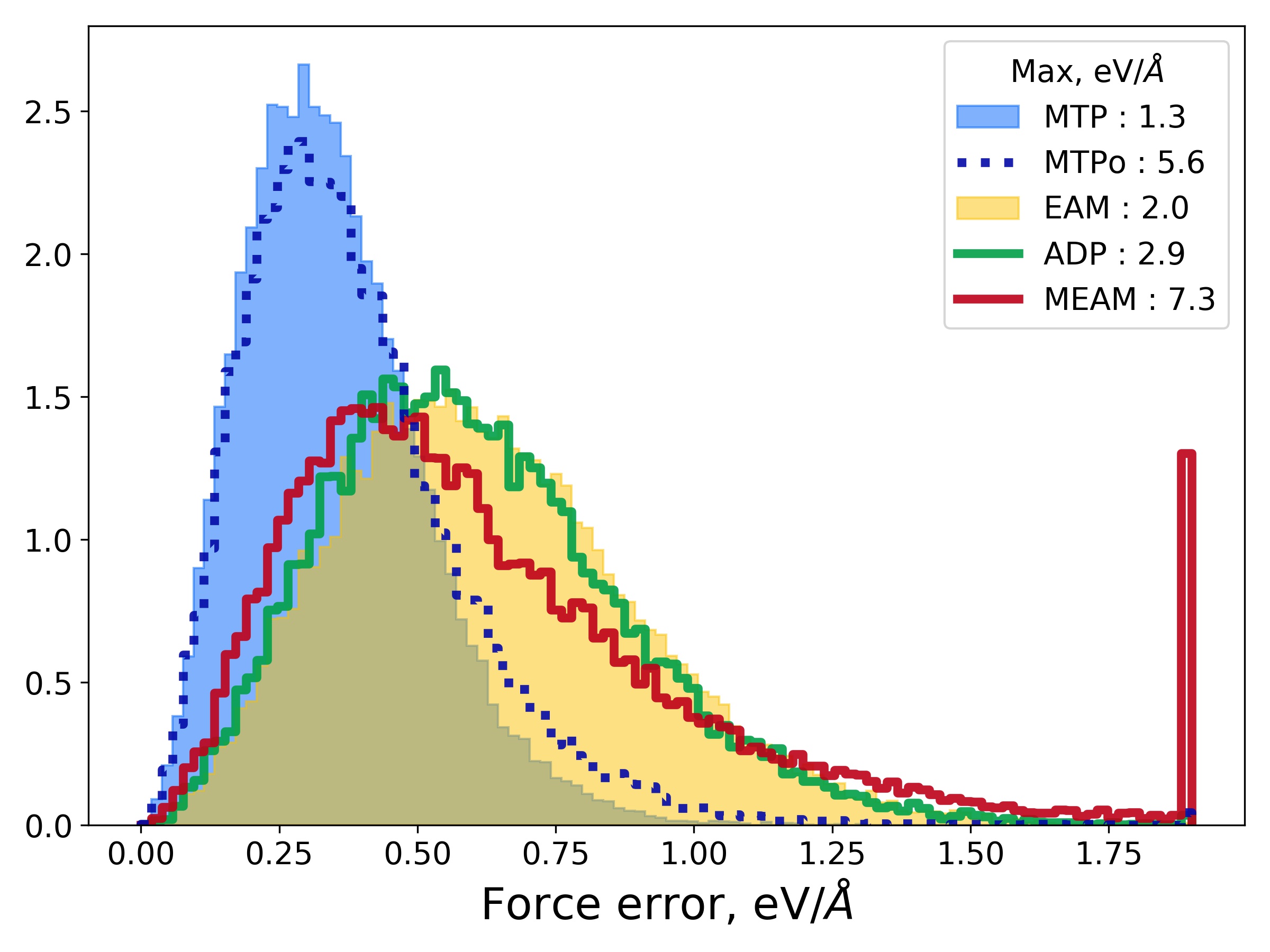}
\put(100, 50){\rotatebox{-90}{\small Force}}
\end{overpic}
\begin{overpic}[width=0.49\textwidth]{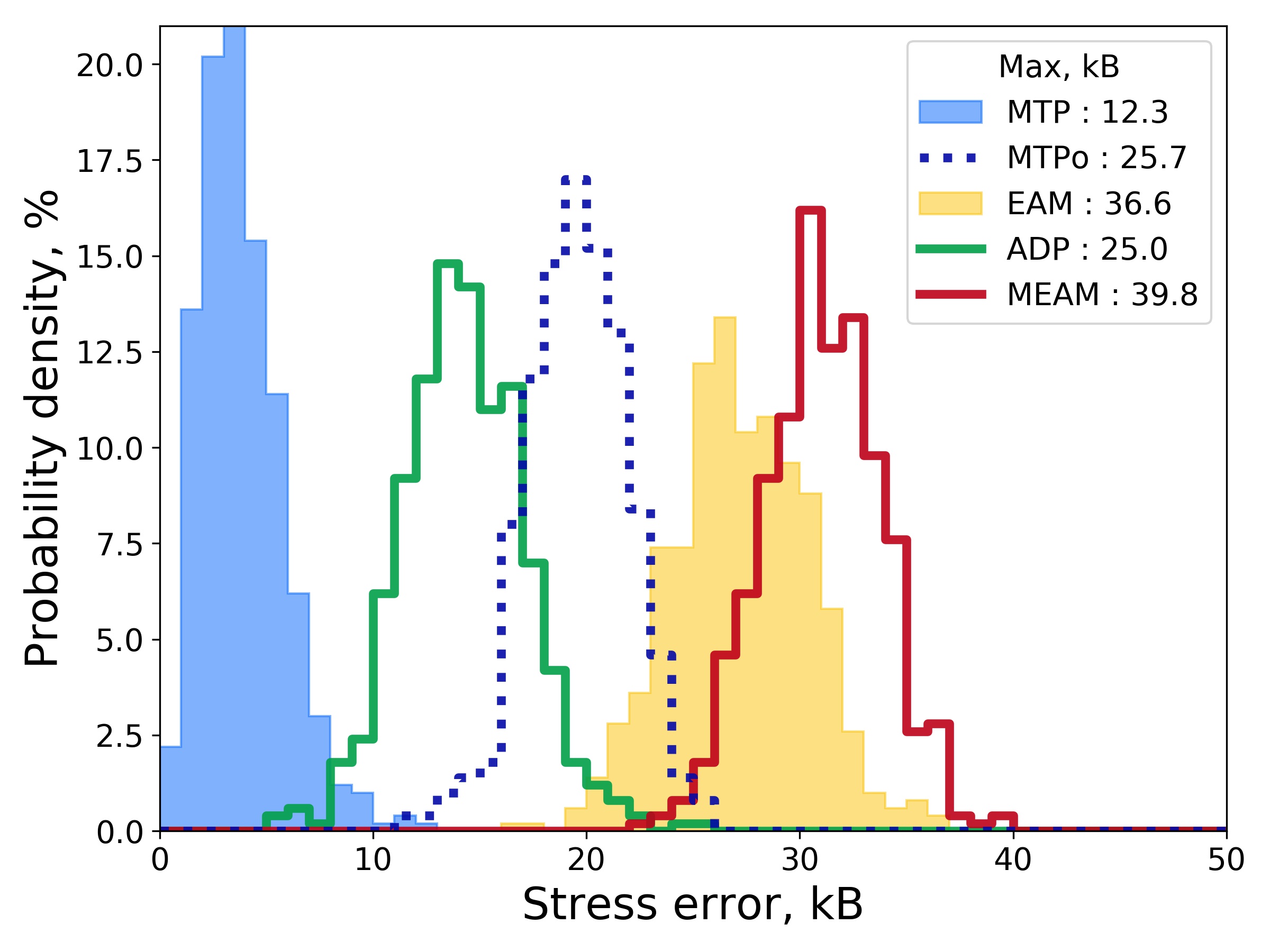}
\end{overpic}
\begin{overpic}[width=0.49\textwidth]{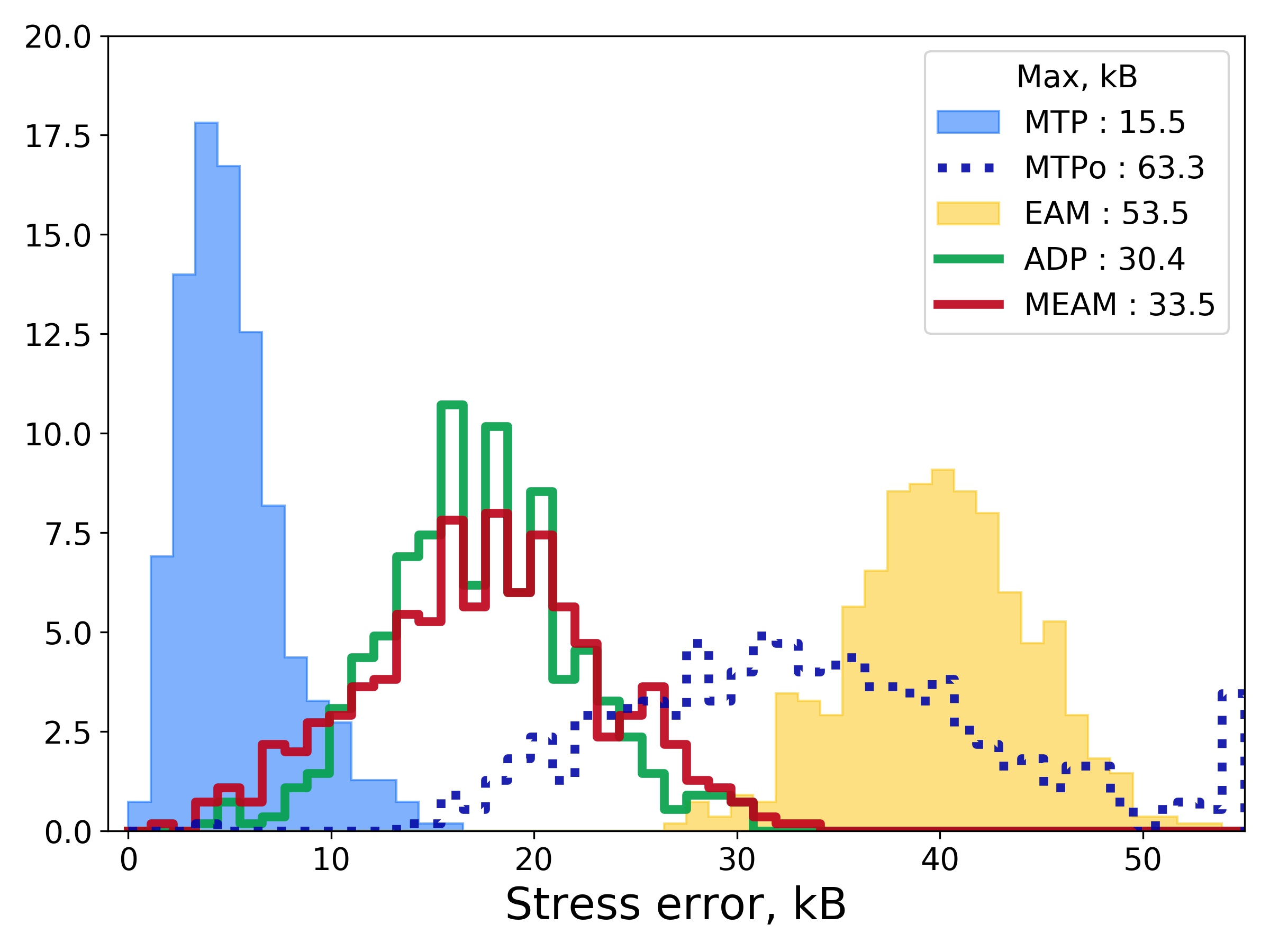}
\put(100, 50){\rotatebox{-90}{\small Stress}}
\end{overpic}

\end{flushleft}

\caption{Distribution of MTP errors for energies (first row), forces (second row) and stresses (third row) in molybdenum. Left column: 1450 K, right: 2600 K.  ``MTPo'' denotes the MTP trained at 2600 K for the left column, and at 1450 K for the right one. Note that the values outside of the plot ranges are placed in the last bin (e.g. look at the MEAM forces at 2600 K).}
\label{fig:EFS_pots_compare}
\end{figure}

MTPs were trained via the active learning on-the-fly algorithm, the length of the corresponding trajectory was 1 ns. For all of the considered potentials, the errors were estimated based on 500 configurations that were randomly selected from equilibrium QMD trajectories of 20 ps.  The training and error estimations were performed at two different temperatures: 1450 K ($0.5\;\mbox{T}_{melt}$) and 2600 K ($0.9 \; \mbox{T}_{melt}$). The resulting error distributions are plotted on Fig.\ref{fig:EFS_pots_compare}.

It should be noted first that the investigated semi-empirical potentials  do not reproduce the absolute energies of configurations. Hence the mean values were subtracted from the distributions of energy errors in order to facilitate the comparison on Fig.\ref{fig:EFS_pots_compare}. The subtracted values are given in the legends. It can be seen from the figure that MTP has both the lowest mean error and the narrowest error distribution at both of the temperatures. 

Interestingly, an increase in the number of adjustable parameters in the sequence EAM-ADP-MEAM does not lead to notable enhancement of accuracy of the potential, at least at 2600 K. At this temperature, MEAM yields a wider distribution of errors than EAM, and it also gives uncertainties larger than 30 meV/atom which are stacked in the last bin. However, the accuracy of MEAM at 1450 K is somewhat better than that of EAM and ADP.

Similar tendencies can be observed on the force error distributions on Fig.\ref{fig:EFS_pots_compare}. MTP is significantly more accurate than MEAM, ADP or EAM at both temperatures. MEAM yields large maximum errors, but the most frequent errors in this case is a bit smaller than that of EAM and ADP. Considering the latter two, EAM yields a smaller maximum error at the high temperature, but a larger mean error at 1450 K.

As expected, stresses are also reliably reproduced by MTP. Noteworthy, the stress errors of both EAM and ADP decrease with temperature, while the MEAM errors, on the contrary, demonstrate a slight increase.

One could justifiably object at this point, that it is not entirely correct to compare potentials that are supposed to work in a wide range of temperatures (EAM, ADP, MEAM) with the one that was fitted at this particular temperature (MTP). To address this concern, the MTP trained at 2600 K was tested with respect to the configurations obtained at 1450 K and vice versa. The results of the tests are plotted on Fig.\ref{fig:EFS_pots_compare} with dotted lines.

Let us denote the potentials trained at 2600 and 1450 K as the ``hot'' and the ``cold'' one, respectively. It can be seen from Fig.3 that the ``hot'' potential works fine at 1450 K, since it yields approximately the same error distributions as at 2600 K. Despite the fact that its accuracy at 1450 K is lower compared to the ``cold'' MTP, it is still notably better than that of EAM, ADP or MEAM.

The performance of the ``cold'' potential at 2600 K is substantially different. It describes energies similarly to EAM or ADP, except for much smaller mean error. Forces are reproduced much better, almost as good as in the case of the ``hot'' MTP. However, a closer examination reveals that the maximum force error of the ``cold'' MTP is four times larger as compared to the ``hot'' potential.

Considering the stress error histograms on Fig.\ref{fig:EFS_pots_compare}, the ``hot''  potential performs better than EAM or MEAM at low temperature. The ``cold'' MTP demonstrates fairly wide error distribution at 2500 K with large maximum error of 63.3 kB, however the value is comparable with that of EAM (53.5  kB). Therefore, it can be concluded that MTP trained at significantly different temperature reproduces stresses similarly to semi-empirical potentials.

In general, it follows from Fig.\ref{fig:EFS_pots_compare} that the ``hot'' MTP employed at the low temperature yields higher mean, but acceptable maximum errors. The ``cold'' potential, on the contrary, yields large maximum errors, when used at the high temperature.

This behavior is likely caused by the features of the selection strategy. It is worth to remind that the employed D-optimality criterion makes MTP to learn very unlike configurations. In other words, MTP selects configurations in order to maximize the phase space volume covered by them. As can be seen from Fig.\ref{fig:EFS_active}, this strategy allows one to reduce maximum errors, but also causes a slight increase in the mean error.

Now consider the two temperatures: high and low. The phase space area covered by an MD trajectory usually increases with temperature. Therefore, configurations selected while training the ``hot'' MTP cover the whole phase space area of the low-temperature trajectory. This explains why the ``hot'' MTP yields similar distributions of energy and force errors at both temperatures.

The ``cold'' potential, on the other hand, has no information about certain areas of the high-temperature phase space, because they were not covered by the low-temperature trajectory. Hence the ``cold'' MTP has to extrapolate, while being employed at the high temperature. This, in turn, leads to large maximum errors.

To summarize, a rule of thumb is that MTPs can be used at a lower temperature than the one they were trained on. However, it is not recommended to employ the potential at higher temperatures, because large maximum errors are expected. If one needs an MTP for higher temperatures, it is much better to use the ``cold'' MTP as a starting point and then to improve it via the active learning on-the-fly algorithm.

\subsection{Application of MTP to diffusion}

It was demonstrated in the previous part of the paper that MTPs are generally more accurate than semi-empirical interatomic potentials. However, all the tests discussed above evaluate only the ability to reproduce the results of DFT calculations. Let us now perform more realistic evaluations by employing MTP to compute vacancy diffusion coefficients in aluminum, molybdenum and silicon.


\subsubsection{Analysis of selection process}

\begin{figure}[h!]
\center 
\includegraphics[width=0.7\textwidth]{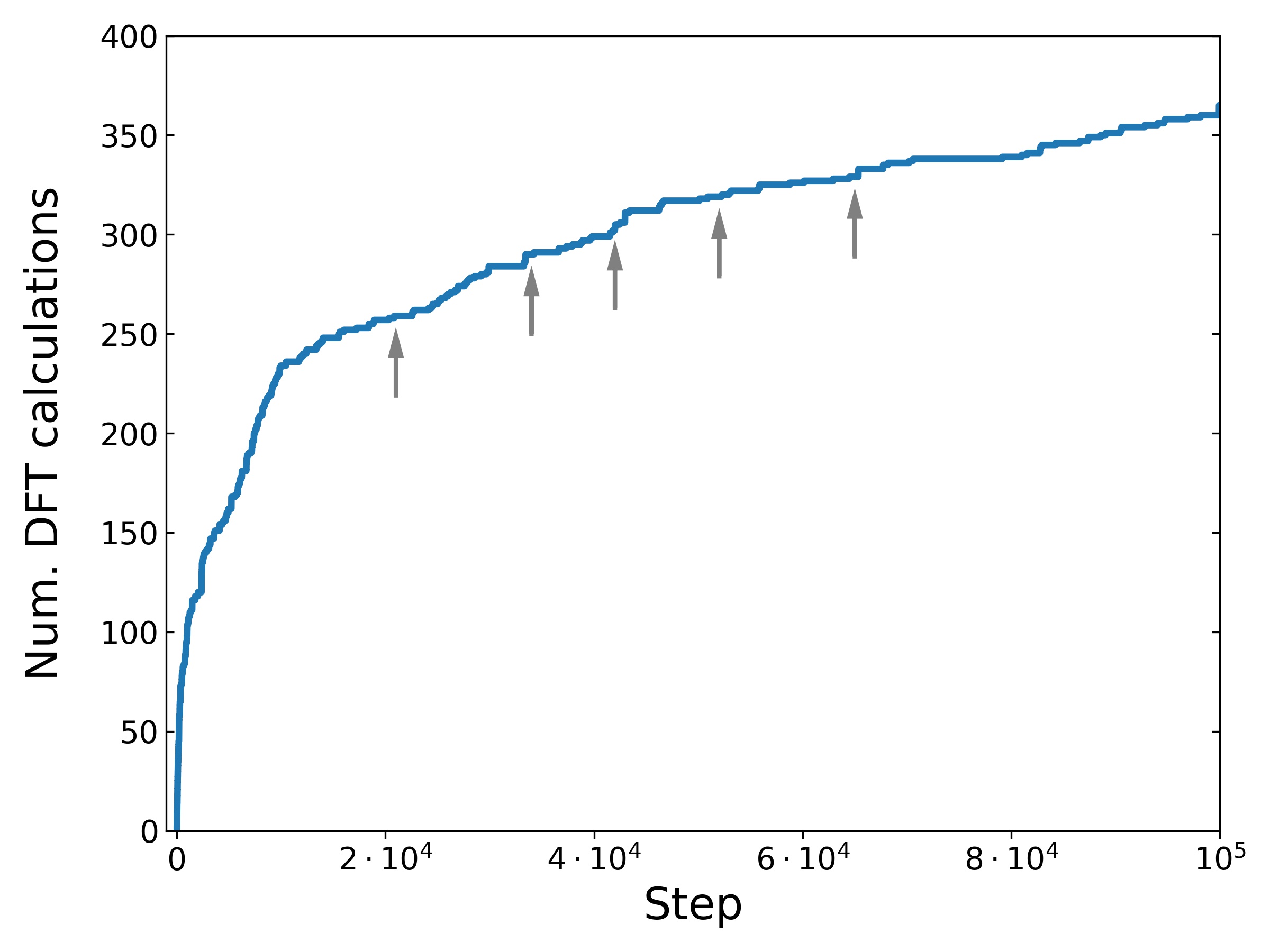}

\caption{The number of DFT calculations as a function of simulation step number during learning on-the-fly of MTP. The results were obtained for Al at $800 \; K$. Arrows indicate the MD time steps, when diffusion jumps occurred.}
\label{fig:lotf_example}
\end{figure}

Diffusion was chosen as a test problem because one may suppose that MTPs and the corresponding learning on-the-fly algorithm are particularly suitable for investigation of rare-event processes. Indeed, an MD trajectory wanders inside a potential basin most of the time in this case. It means that the system remains in the same area of the phase space most of the time. Therefore, MTP can be used to learn this area and replace expensive DFT calculations, at least during the long waiting periods between transition events. Moreover, MTP should recognize the moment when the event eventually occurs and invoke the DFT code. We plotted the number of DFT calls as a function of the simulation time on Fig.\ref{fig:lotf_example} in order to visualize this kind of behavior.

One can clearly see from the figure that the number of DFT calls grows rapidly after the start of the run. Then the curve starts to level out approximately at the step of $10^4$. The flattening of the curve indicates that the potential surface is already well-sampled by the MTP. However, the system continues to explore the phase space, hence new configurations are occasionally selected. 

Often a series of DFT calculations is made in a row, indicating that the trajectory escaped from the known area of the phase space. Such moments can be seen on Fig.\ref{fig:lotf_example} as steps on a flat part of the curve. It is natural to expect that the positions of the steps will correlate with the diffusional jumps, since new areas of the phase space are visited in this case. The jump times are indicated on Fig.\ref{fig:lotf_example} with arrows. It is evident from the plot that the correlation, if present, is not very strong. For example, no steps are visible at the times of the first, the second and the fourth jumps.

One possible explanation of this effect is that the system samples phase space near the dividing surface during unsuccessful jump attempts, hence MTP does not have to learn new configurations when the actual jump occurs. Another possibility is that, even if the learning occurs, only a few configurations could be selected, therefore no step is visible on Fig.\ref{fig:lotf_example}.

\begin{figure}[h!]
\center 
\begin{overpic}[width=0.8\textwidth]{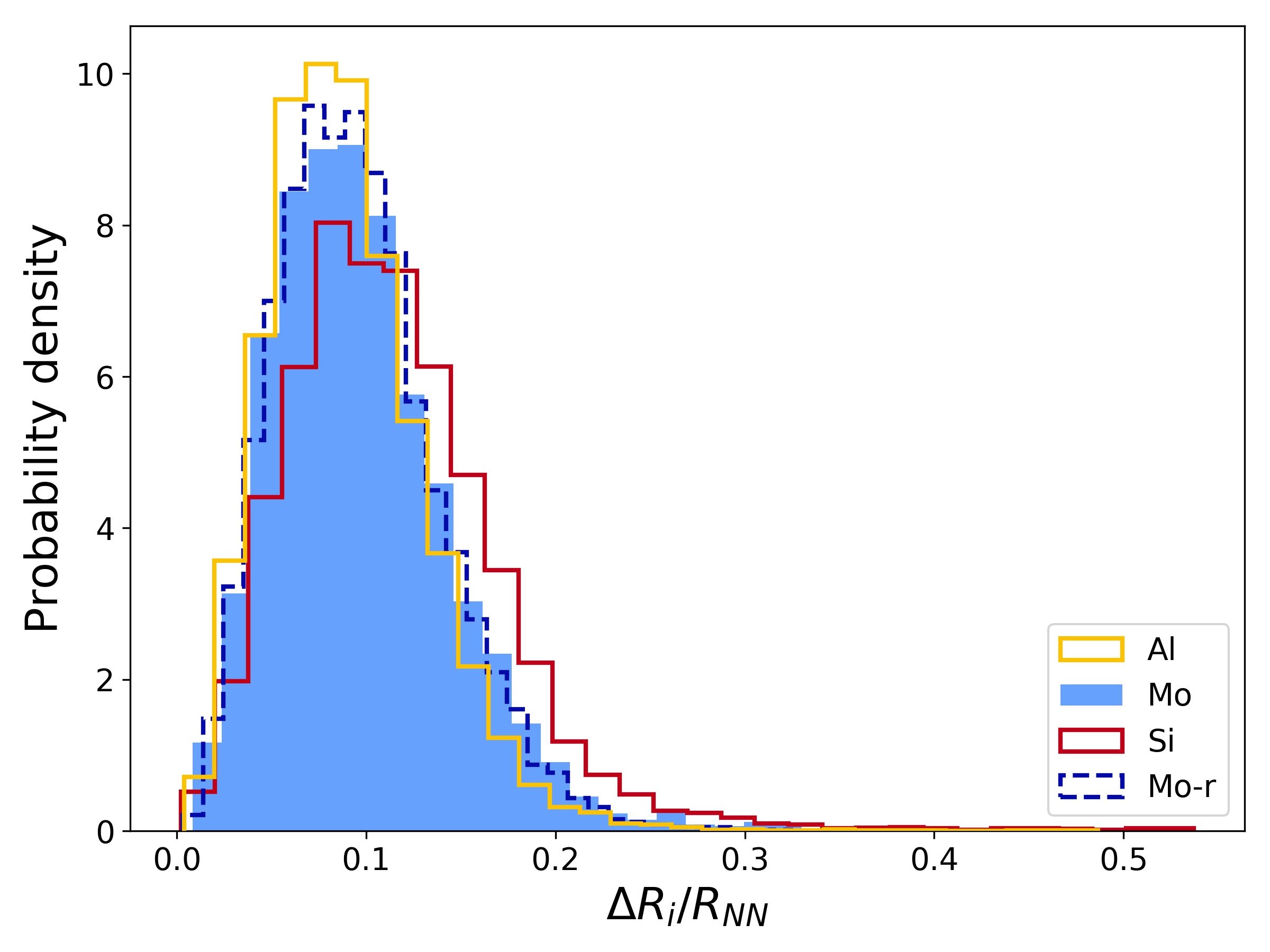}
	\put(39.5, 34.7){\includegraphics[width=0.468\textwidth]{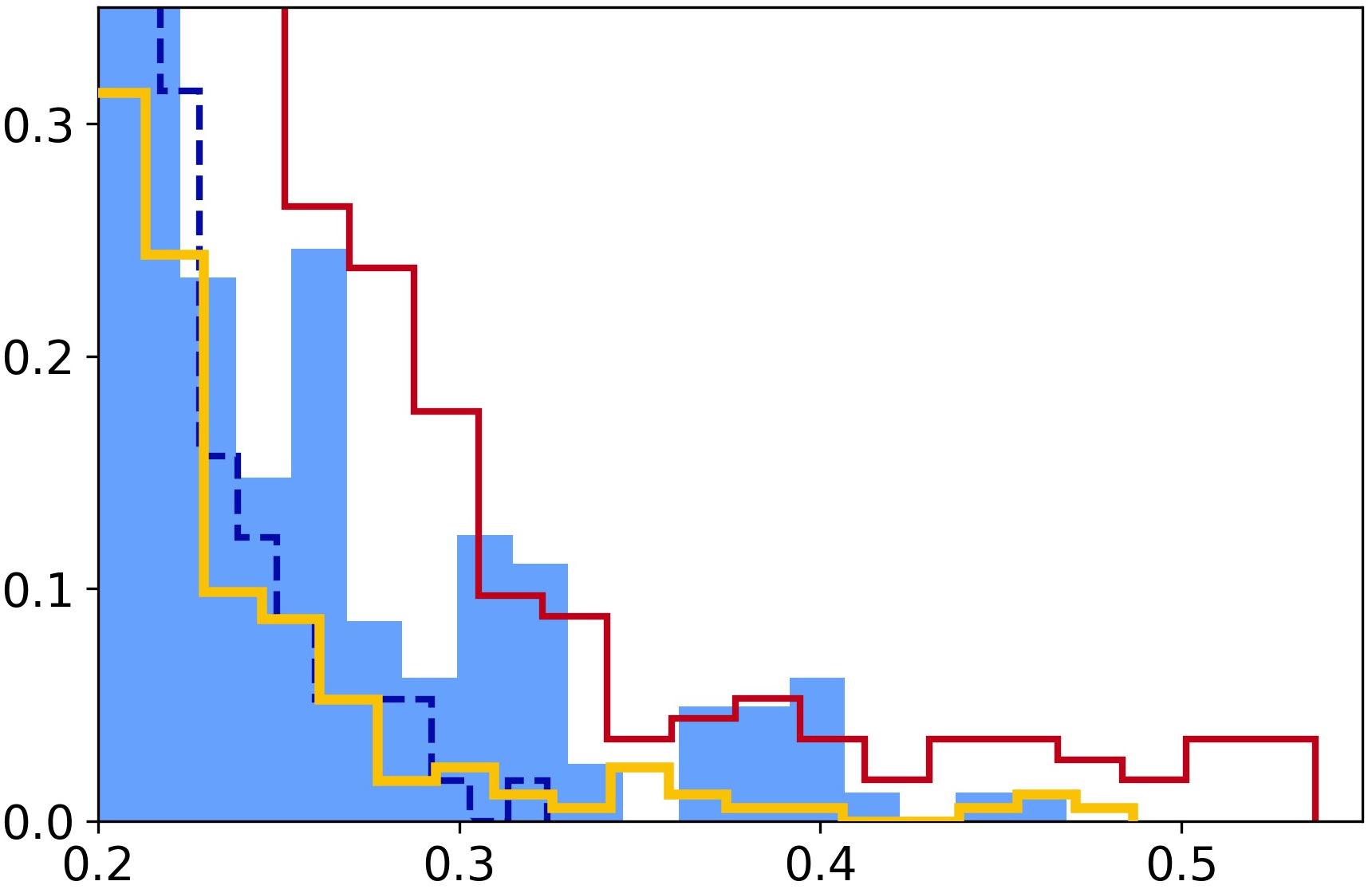}}
\end{overpic}

\caption{Distribution of atomic displacements in the training sets, selected during active learning on-the-fly. All trajectories correspond to the same homologous temperature of $0.75 \; T_m$. The ``Mo-r'' curve corresponds to the randomly selected set of configurations.}
\label{fig:lotf_displ}
\end{figure}

%


It is useful to visualize training sets in order to understand whether configurations near the saddle points are actually selected. Therefore, we considered training sets after active learning on-the-fly, the same as in the Section \ref{compare}. A set of configurations randomly chosen from Mo trajectory was also considered for the purpose of comparison. Then atoms in the configurations were relaxed to their equilibrium positions via conjugate-gradient energy minimization algorithm implemented in the VASP code. Displacement magnitude of an atom during the minimization was subsequently used as a natural measure of its proximity to a saddle point. 

Distributions of the displacements are plotted on Fig.\ref{fig:lotf_displ}. Note that their values are normalized to the nearest-neighbor distance $R_{NN}$  for each of the materials. The employed $R_{NN}$ values are $2.89$, $2.74$ and $2.36 \; \mbox{\AA}$  for Al, Mo and Si, respectively.

As depicted on Fig.\ref{fig:lotf_displ}, the shapes of the distributions for different materials are rather similar. Moreover, the distributions for the randomly (``Mo-r'') and actively (``Mo'') selected sets are also very alike. The latter fact is interesting because active learning selects divergent configurations, therefore one may expect that the distribution for the ``Mo'' set would be shifted to the right with respect to the ``Mo-r''.

This similarity could be due to the fact that the selection algorithm chooses not individual atomic environments, but the whole configurations. Therefore, beside ``interesting'' environments, dozens of other fairly ``usual'' ones are also added to the training set. These considerations suggest that only the ``tales'' of the distributions should be different, and the inset on Fig.\ref{fig:lotf_displ} demonstrates that they indeed are. It can be seen from the plot that the maximum displacements in randomly sampled configurations are less then $0.32 \; R_{NN}$, while the actively selected set include $\Delta R_i$ up to $0.45 \; R_{NN}$.

One can also see from the inset that Al, Mo and Si sets contain dissimilar numbers of largely-displaced atoms. This is likely caused by the fact that the materials have different diffusivities even at close homologous temperatures. Indeed, the Si trajectory contains more then a thousand of the jumps, Mo -- around 500, and only about 10 diffusive hops has happened while the MTP for Al was trained. 

\subsubsection{Calculation of diffusion coefficients}

Let us now proceed to the calculation of diffusion rates in order to evaluate the ability of MTP to reproduce properties of configurations near the dividing surface. As follows from Fig.\ref{fig:EFS_passive}, MTP errors for aluminum are lower than  those for the other materials, therefore it is convenient to consider aluminum first.

\begin{figure}[h!]
\center
\begin{overpic}[width=0.7\textwidth]{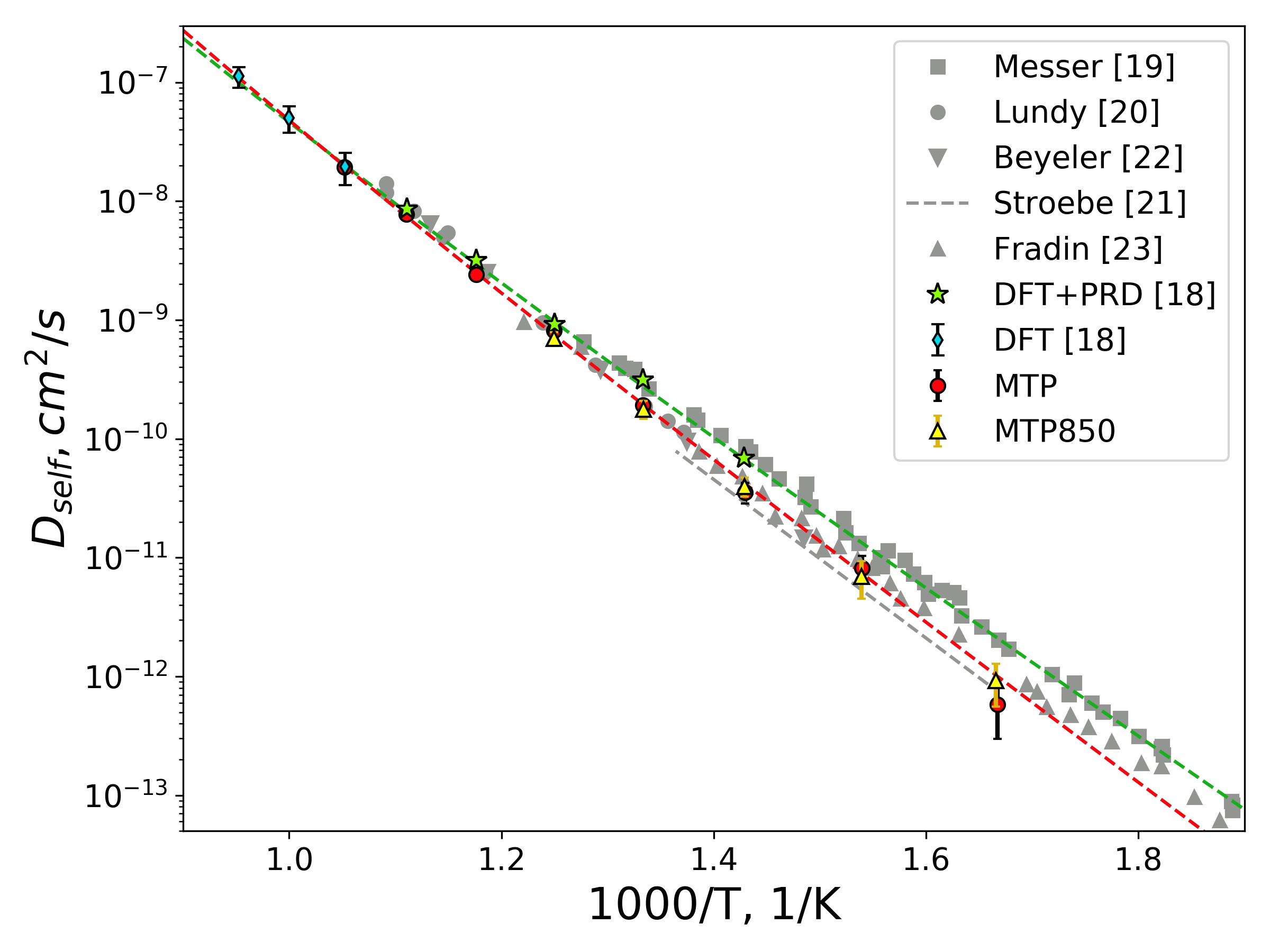}
\end{overpic}
\caption{Coefficients of self-diffusion in aluminum. Experimental data is in grey. Green and red dashed lines correspond to Arrhenius fit of the ``DFT+PRD'' and ``MTP'' series, respectively. The MTP point at 600 K is ignored during the fit, see text for details.}
\label{fig:diff_Al}
\end{figure}

The coefficients of self-diffusion in aluminum were calculated in a standard way as described in the Methods, the results are plotted on Fig.\ref{fig:diff_Al}. The figure also depicts the diffusivities obtained from the conventional (``DFT'') and accelerated (``DFT+PRD'') quantum MD \cite{Novoselov_QMD_PRD}, as well as experimental data (given in grey) \cite{Lundy_Al_diffus, Beyeler_Al_diffus, Stroebe_Al_diffus, Messer_Al_diffus, Fradin_Al_diffus}. 

It can be seen from Fig.\ref{fig:diff_Al} that the results of MTP calculations nicely agree with the ab-initio data at high temperatures, however notable discrepancies are observed at temperatures below 800 K. In particular, diffusivities estimated from MTP simulations at  750 and 700 K are approximately two times lower than the values obtained from PRD-accelerated DFT.

One could speculate that the discrepancies are caused by overestimation of the diffusivities obtained from accelerated DFT, since rather limited statistics is available at low temperatures in this case. Another plausible explanation is that the MTP underestimates diffusion rates at low temperatures due to poor sampling of the phase space near the dividing surface. As mentioned above, only about 10 jumps have happened during active learning of MTP at 700 K. A reasonable way to cope with the problem might be to employ MTP trained at a higher temperature. Such potential is expected to provide more accurate description of infrequent configurations, e.g. during diffusion jumps. On the other hand, it follows from Fig.\ref{fig:EFS_active} that more frequently encountered atomic environments should still be reliably described in this case.

Following these considerations, we employed an MTP trained at 850 K (hereafter MTP850) to calculate diffusion rates at lower temperatures. The results are depicted on Fig.\ref{fig:diff_Al} as the ``MTP850'' series. It is clearly seen from the figure that the obtained diffusivities are very close to the  previously calculated values (``MTP'' series). The only exception is the temperature of 600 K, where MTP850 yields significantly higher diffusion coefficient than the MTP trained at 600 K. However, the discrepancy is not surprising, since no diffusion jumps has happened during active learning on-the-fly of the MTP at 600 K. Therefore the 600 K point of the ``MTP'' series will be disregarded in the analysis below.

The observed correspondence of the ``MTP'' and ``MTP850'' series suggests that the underestimation of diffusion rates with respect to PRD-accelerated DFT is not caused by insufficient sampling of the phase space. Therefore, we are left to conclude that MTP overstates vacancy migration energy for some reason, e.g. due to specifics of the employed basis functions (see Eq.\ref{mtp_repres} for details). Indeed,  Arrhenius fit of  the ``DFT+PRD'' series yields migration energy of 0.55 eV, while the fit of ``MTP'' data gives 0.65 eV. Nevertheless, both of the values are close to 0.6 and 0.57 eV  obtained in \cite{Mattsson2002} and \cite{Mantina2008}, respectively, in the framework of DFT.  Moreover, both MTP and DFT results closely correspond with the experimental data in the entire temperature range considered. 

 

\begin{figure}[h!]
\center
\begin{overpic}[width=0.7\textwidth]{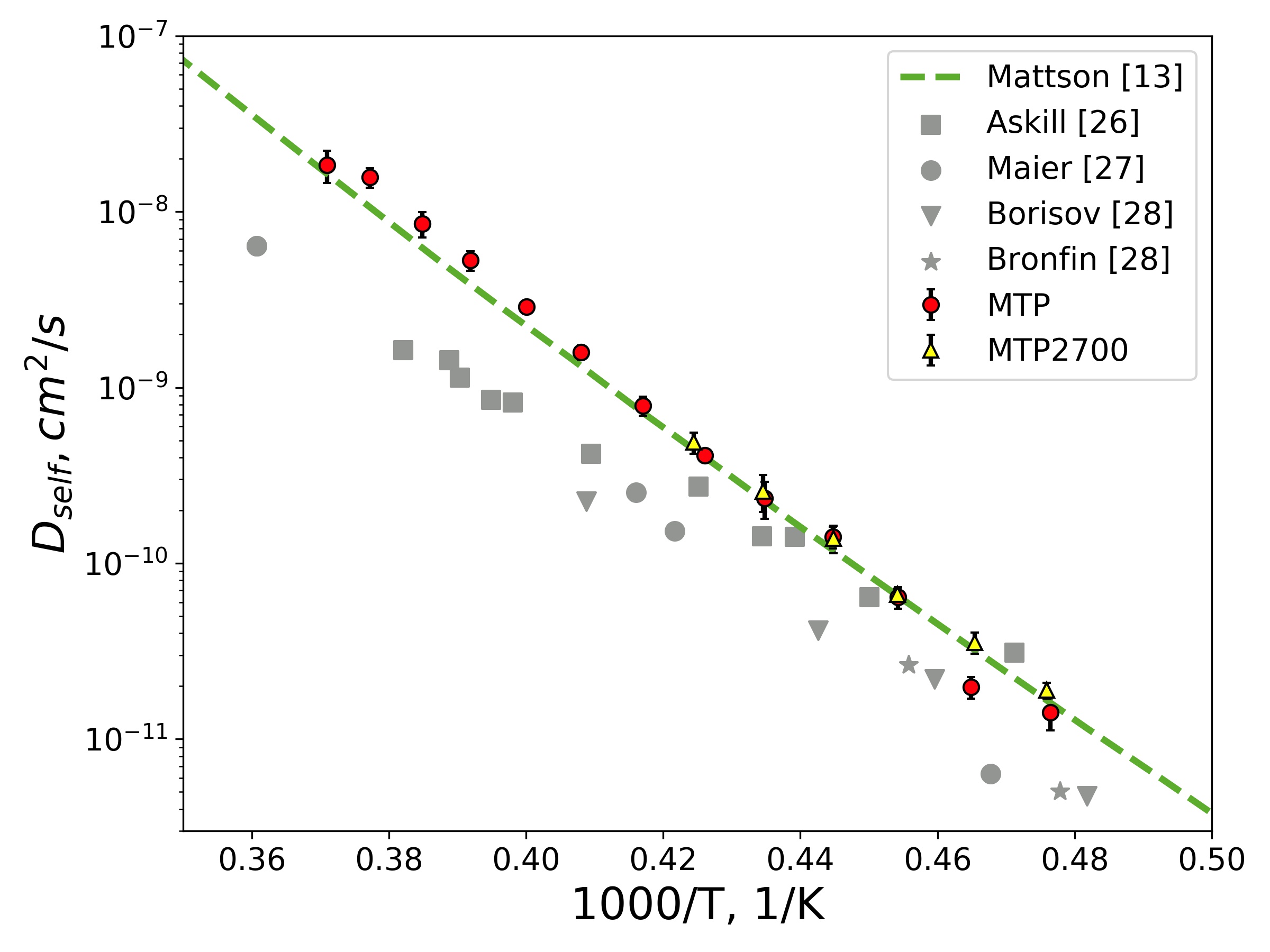}
\end{overpic}
\caption{Coefficients of self-diffusion in molybdenum. Experimental data is in grey. ``MTP2700'' -- the MTP trained at 2700 K is employed to calculate diffusivities at lower temperatures.}
\label{fig:diff_Mo}
\end{figure}

Let us now consider self-diffusion of molybdenum. The values of diffusion coefficients were calculated in the same way as for aluminum, the data is plotted on Fig.\ref{fig:diff_Mo}. The figure also presents the results of DFT calculations (``Mattson'') \cite{Mattsson2009} and experimental data (in grey) \cite{Askill1963, Maier1979, Pergamon_handbook}.

At first glance, the results of MTP calculations match DFT data fairly well. However, a closer look reveals that the MTP points are notably lower than the DFT curve at temperatures below 2350 K. Note that a similar behavior was observed for aluminum.  Therefore, as in the case of Al, an MTP trained at a high temperature (2700 K) was employed to calculate diffusion coefficients at lower temperatures. The results are plotted on Fig.\ref{fig:diff_Mo} as the ``MTP2700'' series. It could be seen from the figure that the high-temperature MTP indeed demonstrates a better agreement with the DFT data compared to the potentials trained at lower temperatures.



\begin{figure}[h!]
\center
\begin{overpic}[width=0.7\textwidth]{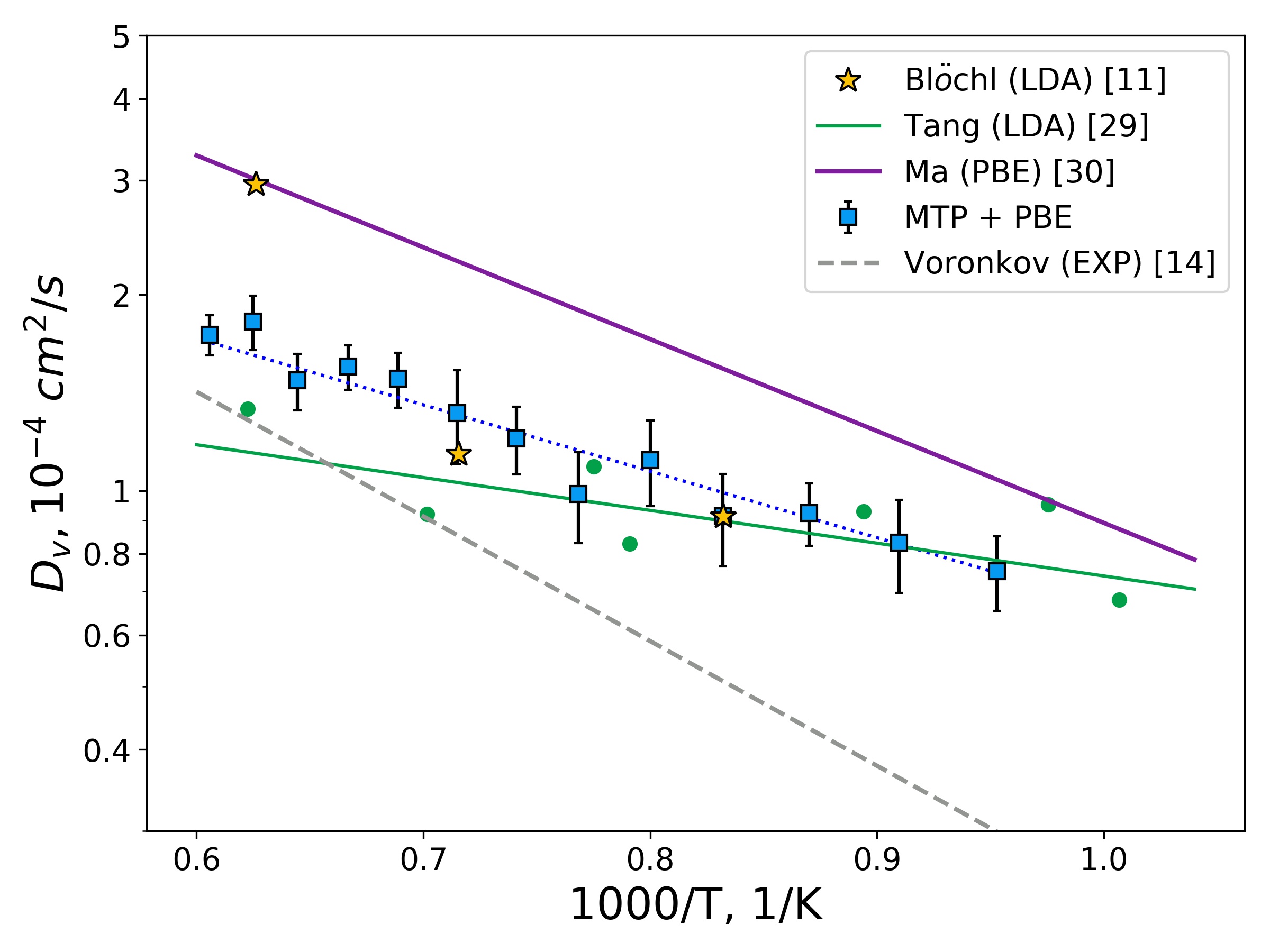}
\end{overpic}
\caption{Vacancy diffusion coefficient in silicon. Error bars correspond to standard deviation of the points.  The data of Koizumi et.al. \cite{Koizumi2011} is not shown on the plot, since it is several orders of magnitude lower than the other values. }
\label{fig:diff_vac_Si}
\end{figure}

Next, we employ MTPs to investigate self-diffusion in silicon. Note that this is a rather complicated process that is still being extensively studied. However, it seems established that, unlike Al and Mo, self-diffusion in silicon occurs through migration of both vacancies and interstitials.

Interstitials are believed to be charged \cite{Ma2010, Ural2001} and dominate self-diffusion at temperatures above 1170 K \cite{Bracht1995, Bracht1998, Shimizu2007}. Their diffusion properties are relatively well-established compared to vacancies, since extremely precise measurements of self-diffusivities are required in the latter case. Discrepancies between estimations of vacancy contribution to self-diffusion \cite{Shimizu2007, Bracht2013, Sudkamp2016} indicate that this task is very challenging from the experimental point of view, see discussion in \cite{Sudkamp2016} for details. Therefore, an application of MTP to this problem could be of interest.

Vacancy diffusion coefficients in silicon were calculated the same way as it was done for aluminum and molybdenum. The results are depicted on Fig.\ref{fig:diff_vac_Si} along with the literature data. It can be seen from the plot that the values obtained with MTPs are in reasonable agreement with the other theoretical result, especially the LDA calculations \cite{Blochl1993, Tang1997}. However, the correspondence with experimental data \cite{Voronkov2006} is rather limited, especially considering the values of migration energy.

\begin{table}[h!]
\begin{tabular}{|c|c|c|c|}
\hline
$D_0, \; cm^2/s$ & $H_m, \; eV$ & Method & Ref.\\
\hline
$2.36 \cdot 10^{-4}$ & $0.1$ & LDA + TBMD & Tang et. al. \cite{Tang1997}\\
\hline
$1.7 \cdot 10^{-6}$ & $0.13$ & LDA + MD & Koizumi et. al. \cite{Koizumi2011}\\
\hline
$2.3 \cdot 10^{-3}$ & $0.28$ & PBE + hTST & Ma and Wang \cite{Ma2010}\\
\hline
$7 \cdot 10^{-4}$ & $0.2$ & PBE + MTP & this work\\
\hline
- & $0.18$ & PBE + NEB & this work\\
\hline
 -  & $0.57$ & HSE06 + NEB & \'{S}piewak et. al. \cite{Spiewak2013}\\
\hline
$2. \cdot 10^{-3}$ & $0.38$ & Experiment & Voronkov et. al. \cite{Voronkov2006}\\
\hline
$1.2 \cdot 10^{-3}$ & $0.45$ & Experiment & Watkins \cite{Watkins2008}\\
\hline
\end{tabular}
\caption{Vacancy diffusion parameters obtained by different methods. TBMD -- tight binding MD; hTST -- harmonic transition state theory. Note that the $D_0$ values published in \cite{Koizumi2011} and \cite{Tang1997} were divided by $0.5$, since correlation effects were not considered in the original papers.}
\label{tab:diffus_table}
\end{table}

The Arrhenius parameters, namely the pre-exponential factors and migration energies, are given in Tab.\ref{tab:diffus_table} in order to facilitate further comparison. One can see from the table that migration energies are indeed severely underestimated in LDA calculations. The generalized gradient approximation in the PBE form yields significantly higher migration energies, but they are still twice as low as compared to the experimental values. These disadvantages of LDA and GGA-PBE with respect to silicon are well-known and discussed in the literature \cite{Spiewak2013, Batista2006} where it is suggested that hybrid exchange-correlation functionals, e.g. HSE06, should be employed in order to obtain more reliable results.


It is worth to remind that several hundreds of DFT calculations are required in order to train MTP, even for a single temperature. This task is hardly feasible for hybrid functionals, since calculation of the Hartree-Fock exact exchange term is exceptionally demanding from the computational point of view. Therefore, the PBE functionals were employed in this work. Indeed, our aim was not to calculate accurate vacancy diffusion rates in Si, but rather to demonstrate the applicability of MTP to the investigation of diffusion in non-metallic materials. Therefore, we calculated the migration energy via the nudged elastic band (NEB) method in order to validate the MTP results. It can be seen from Tab.\ref{tab:diffus_table} that NEB yields $H_m = 0.18 \; \mbox{eV}$, which is close to $0.2 \; \mbox{eV}$ obtained from MTP calculations.


Note that Ma and Wang \cite{Ma2010} obtained considerably different NEB results. They report the migration energy that is 0.1 eV higher than the value obtained in this work. Additional analysis revealed that the discrepancy is caused by the difference in the employed lattice constants. We adopted the experimental lattice parameter of $5.445 \; \mbox{\AA}$ \cite{Si_expans}, while the theoretical value of  $5.475 \; \mbox{\AA}$ was employed in \cite{Ma2010}. Using the latter value, we also obtained the 0.28 eV barrier.

\begin{figure}[h!]
\center
\begin{overpic}[width=0.7\textwidth]{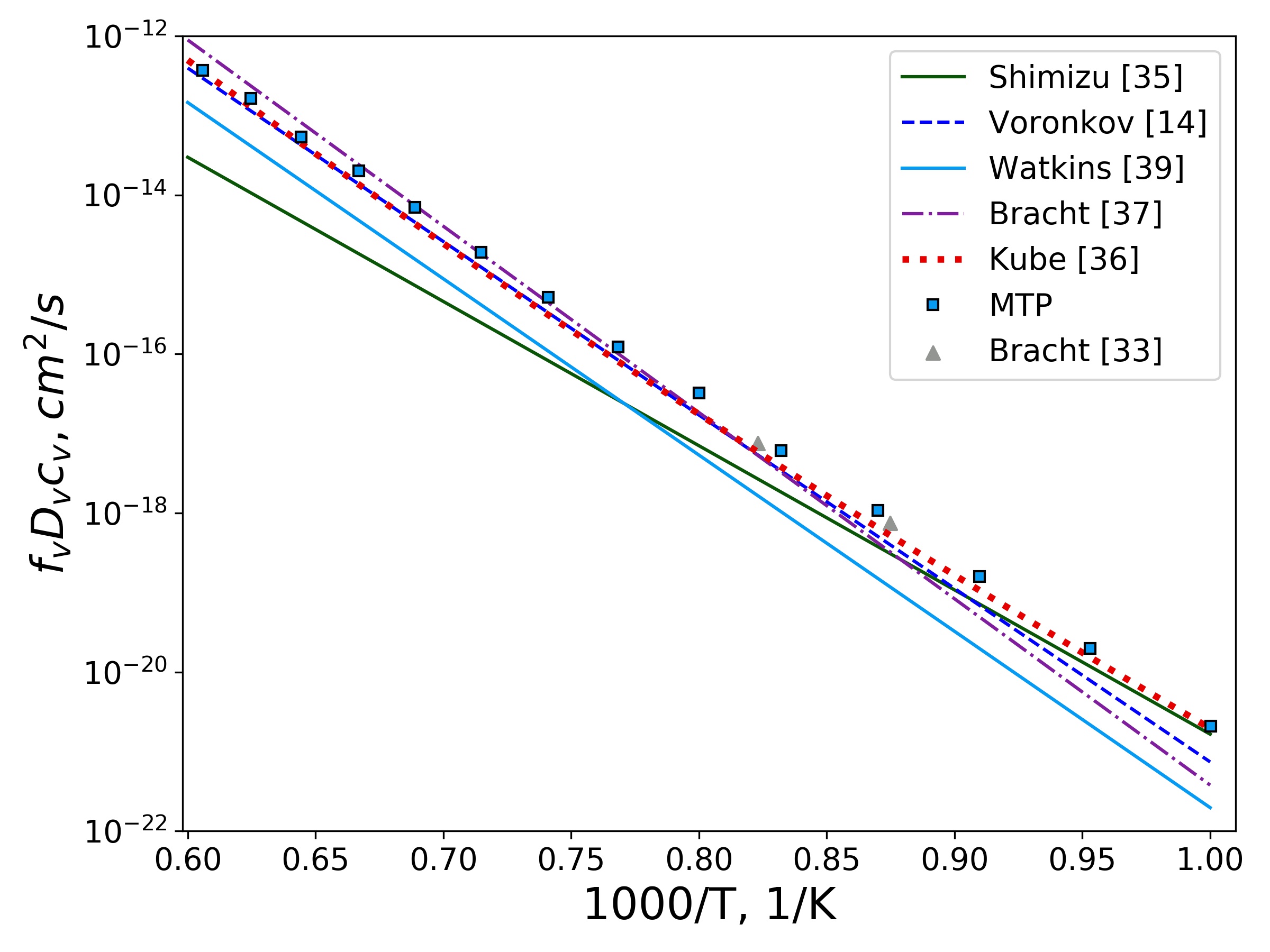}
\end{overpic}
\caption{Vacancy contribution to self-diffusion in silicon. Error bars are smaller than the point sizes.}
\label{fig:diff_Si}
\end{figure}

Now, when we are sure that MTP reliably reproduces DFT data, the vacancy diffusion coefficients can be used to calculate corresponding contribution to self-diffusion. This was done in a standard way, as discussed in the Methods section. The vacancy formation parameters were adopted from the experimental work \cite{Voronkov2006}. The results are plotted on Fig.\ref{fig:diff_Si} along with the experimental data. One can see from the figure that, despite the significantly different migration energy, the calculated values are still in a reasonable agreement with experiment, at least in the temperature range where the experiments are usually performed.

\section{Conclusion}

A recently proposed class of machine-learning interatomic potentials --- Moment tensor potentials --- was investigated in this work. It was demonstrated that MTP provides accurate description of energies, forces and stresses with respect to the DFT calculations.

 An important feature of MTP is the ability to actively select configurations and train the potential on-the-fly. This feature allows one to automatically parametrize MTP for specific conditions (e.g. temperature and pressure) and thus significantly decrease the amount of required computational resources without substantial loss of accuracy.  This makes MTP a very practical and powerful tool for modeling of materials at atomic scale.

In this work, we employed MTP to calculate vacancy diffusivities in aluminum, molybdenum and silicon. The obtained diffusion coefficients and migration energies are in a good agreement with DFT calculations in all the cases. This is indeed impressive, since the potentials were not specifically fitted to reproduce migration barriers.


\section*{Acknowledgements}
The work was supported by the Russian Science Foundation (grant number 18-13-00479).

\section*{Data availability}
The raw data required to reproduce these findings are available to download from \cite{raw_data}. The processed data required to reproduce these findings are available to download from  \cite{processed_data}.


\bibliography{MLIP_references}

\newpage

\end{document}